\documentclass[twocolumn]{aastex631}

\usepackage{amsmath, amscd, amstext, amsbsy, amsopn, amsthm, amsxtra, upref}
\usepackage{xcolor}
\usepackage{lipsum}

\let\tablenum\relax
\usepackage{siunitx}

\begin{document}
\DeclareSIUnit{\angstrom}{\text{\normalfont\AA}}
\title{Modeling the formation and evolution of solar wind microstreams: from coronal plumes to propagating Alfvénic velocity spikes}

\author[0000-0002-1711-1802]{Bahaeddine Gannouni}
\affiliation{IRAP, Université Toulouse III - Paul Sabatier, CNRS, CNES, Toulouse, France}

\author[0000-0002-2916-3837]{Victor Réville}
\affiliation{IRAP, Université Toulouse III - Paul Sabatier, CNRS, CNES, Toulouse, France}

\author[0000-0003-4039-5767]{Alexis P. Rouillard}
\affiliation{IRAP, Université Toulouse III - Paul Sabatier, CNRS, CNES, Toulouse, France}



\begin{abstract}

We investigate the origin of mesoscale structures in the solar wind called microstreams, defined as enhancements in the solar wind speed and temperature that last several hours. They were first clearly detected in Helios and Ulysses solar wind data and are now omnipresent in the "young" solar wind measured by the Parker Solar Probe and Solar Orbiter. These recent data reveal that microstreams transport a profusion of Alfvénic perturbations in the form of velocity spikes and magnetic switchbacks. In this study, we use a very-high-resolution 2.5D MHD model of the corona and the solar wind to simulate the emergence of magnetic bipoles interacting with the preexisting ambient corona and the creation of jets that become microstreams propagating in the solar wind. Our high-resolution simulations reach sufficiently high Lundquist numbers to capture the tearing mode instability that develops in the reconnection region and produces plasmoids released with the jet into the solar wind. Our domain runs from the lower corona to 20 $R_{\odot}$, which allows us to track the formation process of plasmoids and their evolution into Alfvénic velocity spikes. We obtain perturbed solar wind flows lasting several hours with velocity spikes occurring at characteristic periodicities of about 19 minutes. We retrieve several properties of microstreams measured in the pristine solar wind by the Parker Solar Probe, namely an increase in wind velocity of about 100 km/s during a stream’s passage together with superposed velocity spikes of also about 100 km/s released into the solar wind.

\end{abstract}



\section{Introduction} \label{sec:intro}


White-light images of total solar eclipses and coronagraphs reveal fine ray-like structures emanating from polar coronal holes. These "plumes" extend outward from the base of the corona and are observed in white-light and extreme-ultraviolet (EUV) images. They are most commonly found in polar coronal holes, but can also be observed in equatorial coronal holes \citep{1996ApJ...464L..95W}. Plasma and magnetic field data obtained by the two Helios solar probes showed that the scales of these rays are preserved in the evolving interplanetary high-speed solar wind measured close to the Sun. Further studies based on in situ measurements made in the polar solar wind by the Ulysses mission confirmed this result and identified ‘microstreams’ in the form of velocity fluctuations of ±40 km s$^{-1}$, higher kinetic temperatures, slightly higher proton fluxes \citep{1995JGR...10023389N}. \citet{2008ApJ...682L.137R} showed that X-ray jets are precursors of polar plumes and in some cases cause brightenings of plumes. Microstreams could therefore be the interplanetary manifestation of X-ray jets released during the formation of a plume inside a coronal hole  \citep{2012ApJ...750...50N}. The aim of the present study is to investigate, through high-resolution magneto-hydrodynamic simulations, the mechanisms driving the formation and evolution of plumes and microstreams and their dynamic properties discussed in the next paragraphs.\\

Plumes are typically hazy and are routinely detected in the EUV wavelengths of $\SI{171}{\angstrom}$ and $\SI{193}{\angstrom}$ \citep{2014ApJ...787..118R}. It has been debated as to whether coronal plumes or interplume regions may be the source regions of the fast solar wind \citep{2011A&ARv..19...35W,2015LRSP...12....7P}. Plumes appear to form after magnetic bipoles erupt in the open magnetic field of coronal holes. According to \citet{2001ApJ...560..490D}, plumes extend away from photospheric flux concentrations and can last from hours to several weeks, reaching lengths of about 30 solar radii (R$_\odot$). Plumelets, which are small features within plumes, often exhibit intensity fluctuations on shorter time scales than the overall plume \citep{1997SoPh..175..393D,2007ApJ...661..532D,2021ApJ...907....1U}. Data from STEREO/EUVI images show that these fluctuations, known as propagating disturbances, can have periods ranging from $5$ to $30$ minutes \citep{2010A&A...510L...2M,2011ApJ...736..130T}. The formation of a plume is typically preceded by recurrent jets that emerge from random flux emergence and cancellation, and the plume itself goes through phases of brightening and decay, during which subplumes may be visible \citep{2014ApJ...787..118R}.
The emergence of magnetic flux in the dominant polarity of coronal holes plays an essential role in the heating and the outflow of plasma, and EUV brightening \citep{Panesar_2018,Panesar_2019}. This is likely due to an interchange reconnection process, which takes place when emerging loop systems encounter an open background magnetic field \citep{2002ESASP.505..105V}. It is also an efficient means of releasing plasma that is otherwise confined to closed field regions into the heliosphere and perhaps contributes to the mass flow of fast and slow solar winds emerging from coronal holes \citep{1996Sci...271..464W}.

Interchange reconnection has also been one of the suggested mechanisms for the formation of magnetic switchbacks and velocity spikes measured ubiquitously by Parker Solar Probe (PSP) in the nascent solar wind \citep{Bale_2019,2019Natur.576..228K}. Switchbacks are characterized by large-amplitude Alfvénic fluctuations that propagate away from the Sun, with an extensive range of magnetic deflection angles from a few degrees to a full inversion \citep{2022A&A...663A.109F}. The origins of these features are still under debate, particularly whether they are generated locally in the solar wind or in the lower corona. For instance, the work of \citet{2021ApJ...909...95S} and \citet{Squire_2020,2021ApJ...915...52S}, suggests that switchbacks may be generated locally in the solar wind through processes involving velocity shears or turbulent flows. On the other hand, other studies, such as the work of \citet{2020ApJ...894L...4F} and \citet{2021A&A...650A...2D}, propose that switchbacks are formed through interchange reconnection in the lower corona \citep{https://doi.org/10.1029/2003JA010274,2005ApJ...626..563F,2020ApJ...894L...4F}.\\

Switchbacks and velocity spikes come in bursts or patches whose spatial and time scales are comparable to those of microstreams \citep{Bale2021,2021ApJ...919...96F}. These patches of disturbances are particularly intense in streamer flows, but are also very clear in solar wind flows originating from deep inside coronal holes \citep{2020ApJS..246...37R,2021ApJ...919...96F}. Statistical analysis of these patches of switchbacks/velocity spikes \citep{2021ApJ...919...96F} as well as the analysis of solar wind composition \citep{Bale2021} point towards an origin of these patches in sudden energy releases at the boundary of supergranules. Since photospheric transport processes force an accumulation of magnetic elements (loops and open fields) near the boundaries of granules and supergranules, interchange magnetic reconnection could occur frequently in these regions. Moreover, in the study of \citet{Shi2022}, the analysis of PSP co-rotation periods revealed temporal signatures in addition of spatial structure associated with switchback patches. Disentangling spatial and temporal scales in the data and identifying the corresponding processes at the solar surface will be key to evaluating the idea that switchbacks have indeed a solar origin. \\

Several studies have recently appeared in the literature that make the tentative association between microstreams and plumes and individual switchbacks with the jetlets \citep{2023ApJ...945...28R, 2023arXiv230506914K, Hou_2023}. In this paper, we investigate the idea that interchange reconnection can arise from the emergence of magnetic flux that contributes to the formation of coronal plumes. In particular, we wish to study the effect of the rate and amplitude of flux emergence on the formation of coronal plumes, interchange reconnection and the structure of the resulting jets. An association between plasmoids formed during reconnection and switchbacks was proposed by \citep{2021A&A...650A...2D} using kinetic simulations \citep{2023Natur.618..252B}. Recent advanced MHD simulations have also looked into the evolution of reconnection outflows that become in 2.5-D compressible Alfvén waves \citep{2018RAA....18...45Z,He2021} and in 3-D torsional Alfvén waves \citep{2009ApJ...691...61P,2022ApJ...941L..29W} escaping the solar corona. In the present paper, we examine whether the magnetic islands produced through the tearing-mode instability in the reconnection layer could be the source of individual velocity spikes and switchbacks. In order to capture the development of the tearing-mode instability in the solar corona at the adequate time and spatial scales we limit our study to 2.5-D MHD but extend the domain out to 20 Rs. This allows us to simulate the lifetime of an entire microstream and the associated release of multiple microjets to reproduce the form of microstreams measured in situ by PSP.

The paper is organized as follows. In Section \ref{sec:simu}, we describe the numerical model. In Section \ref{sec:Res}, we show the results for the main simulation setup, along with some discussion of the results. In Section \ref{sec:Disc}, we conclude this study and comment on possible future work.
\newpage
\section{Simulation Details} 
\label{sec:simu}

\subsection{Model Description}

\begin{figure*}
    \centering
    \includegraphics[width=0.8\textwidth]{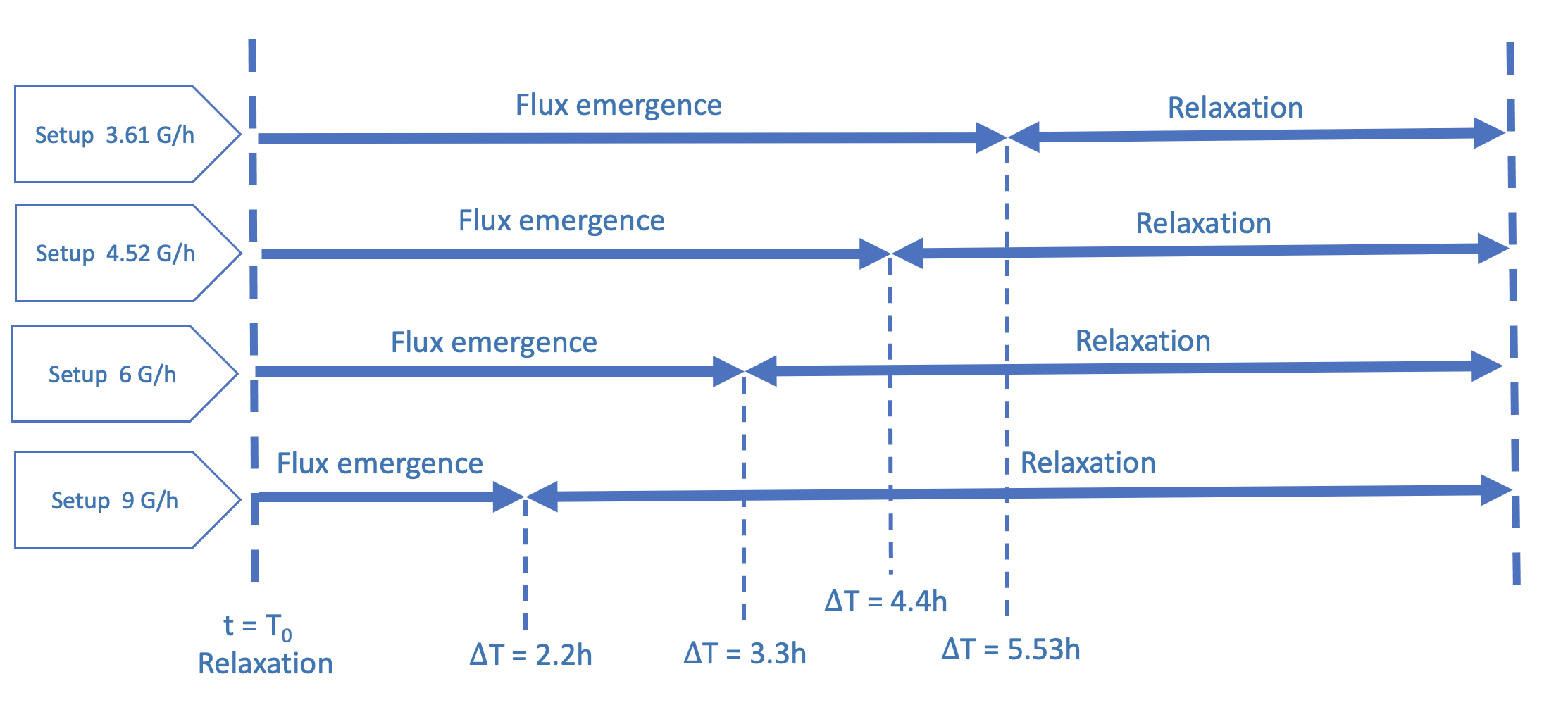}
    \caption{Figure illustrating various simulated setups of flux emergence evolution over time.}
    \label{fig:figsetup}
\end{figure*}

We investigate the dynamics of magnetic reconnection in a solar corona undergoing flux emergence. Flux emergence is a process by which new magnetic field lines emerge from the solar surface and enter the corona, leading to the formation of current sheets \citep[see][]{Shibata1989, Shibata1992a}. One type of reconnection process that has received significant attention is the tearing instability, which occurs when small perturbations in a current sheet grow and lead to the break-up of the sheet into magnetic islands, or “plasmoids”. The tearing mode allows reaching fast reconnection, provided that the Lundquist number $S=L v_A /\eta$ is high enough (L is the length of current sheet, $v_A$ Alfvén velocity, $\eta$ magnetic diffusivity), typically $10^4$ in 2D configurations \citep{Biskamp1986, 2007PhPl...14j0703L,2009PhPl...16k2102B}. A particularly interesting dynamics occurs when a current sheet system is thinning. Linear theory shows that the tearing mode is then triggered as soon as the sheet aspect ratio reaches $\sim S^{-1/3}$ \citep{2014ApJ...780L..19P, Reville2020ApJL}. In our case, as the bipolar flux emerges, a current sheet is formed at the contact of the opposite polarities and interchange reconnection occurs. The current layer is then expected to develop magnetic islands or plasmoids.

To study this process, we solve the 2.5D compressible resistive MHD equations, using the PLUTO code \citep{2007ApJS..170..228M}, a finite-volume shock-capturing code. We employ a second-order Runge–Kutta method to calculate the time step and a fourth-order spatial scheme provided by a parabolic reconstruction. We use a  Harten-Lax-van Leer discontinuities HLLD solver \citep{2005JCoPh.208..315M}. The solenoidal constraint on the magnetic field is ensured through the constrained transport method \citep{BalsaraSpicer1999}.

The equations can be written as follows:

\begin{equation}
\begin{gathered}
\frac{\partial}{\partial t} \rho+\nabla \cdot \rho \boldsymbol{v}=0 
\end{gathered}
\end{equation}
\begin{equation}
\begin{gathered}
\frac{\partial}{\partial t} \rho \boldsymbol{v}+\nabla \cdot(\rho \boldsymbol{v} \boldsymbol{v}-\boldsymbol{B} \boldsymbol{B}+\boldsymbol{I} p)=-\rho \nabla \Phi 
\end{gathered}
\end{equation}
\begin{equation}
\begin{gathered}
\frac{\partial}{\partial t}(E+\rho \Phi)+\nabla \cdot[(E+p+\rho \Phi) \boldsymbol{v} \\-\boldsymbol{B}(\boldsymbol{v}  \cdot \boldsymbol{B}) +(\eta \cdot \boldsymbol{J}) \times \boldsymbol{B}]=Q_h-Q_c-Q_r
\end{gathered}
\end{equation}
\begin{equation}
\begin{gathered}
\frac{\partial}{\partial t} \boldsymbol{B}+\nabla \cdot(\boldsymbol{v} \boldsymbol{B}-\boldsymbol{B} \boldsymbol{v})-\eta \nabla^2 \boldsymbol{B}=0
\end{gathered}
\end{equation}
where $E \equiv \rho e+\rho v^2 / 2+B^2 / 2$ is the background flow energy, $\boldsymbol{B}$ is the magnetic field, $\rho$ is the mass density, $\boldsymbol{v}$ is the velocity field, $p=p_{\text {th }}+B^2 / 2$ is the total pressure (thermal and magnetic), $I$ is the identity matrix, $\boldsymbol{J}= \nabla  \times \boldsymbol{B}$ is  the electric current  , $e=\frac{p_{\mathrm{th}}}{(\gamma-1)\rho}$ is the specific internal energy density and $\eta$ is the magnetic diffusivity. Finally, $\gamma = 5/3$ is the ratio of specific heats, and the terms $Q_*$ represent the terms of volumetric energy gain and loss, heating, thermal conduction, and radiative losses. The system is solved in spherical coordinates (r, $\theta$), and the gravity potential,
\begin{equation}
 \Phi=-\frac{GM_{\odot}}{r}
\end{equation}

The heating term is defined as:
\begin{equation}
Q_h=F_h / H\left(\frac{R_{\odot}}{r}\right)^2 \exp \left(-\frac{r-R_{\odot}}{H}\right)
\end{equation}
The energy flux from the photosphere, denoted as $F_h$, has a value of $1.5 \times 10^5$ erg $\mathrm{cm}^{-2} \mathrm{~s}^{-1}$ 
\citep{1988ApJ...325..442W} and $H \sim 1 R_{\odot}$.

The thermal conduction is written as
\begin{equation}
Q_c=\nabla \cdot\left(\alpha \boldsymbol{q}_s+(1-\alpha) \boldsymbol{q}_p\right)
\end{equation}
where $q_s=-\kappa_0 T^{5 / 2} \nabla T$ is the Spitzer-Härm thermal conduction with $\kappa_0=9 \times 10^{-7} \mathrm{cgs}$, and $\boldsymbol{q}_p=3 / 2 p_{\mathrm{th}} \boldsymbol{v}_e$ is the electron collisionless heat flux described in \citep{1986JGR....91.4111H}. The coefficient $\alpha=1 /\left(1+\left(r-R_{\odot}\right)^4 /\left(r_{\text {coll }}-R_{\odot}\right)^4\right)$ creates a smooth transition between the two regimes at a characteristic height of $r_{\text {coll }}=5 R_{\odot}$.
We have used an optically thin radiation cooling prescription,
\begin{equation}
Q_r=n^2 \Lambda(T)
\end{equation}
with $n$ the electron density and $T$ the electron temperature. $\Lambda(T)$ follows the prescription of \citet{1986ApJ...308..975A}.

In the PLUTO code, all quantities are expressed in dimensionless units derived from physical quantities divided by normalization units, appropriate for solar wind conditions. We consider unit length $L_0=1 \mathrm{R_\odot} \sim 6.9570 \times 10^5$ km, unit density $\rho_0=1.67 \times 10^{-12} \mathrm{~kg} \mathrm{~m}^{-3}$. The velocities are normalized to the keplerian velocity $v_0 \sim 437$ km/s. The characteristic magnetic field and unit time are $B_0=\sqrt{4 \pi \rho_0 v_0^2} \sim 2$ G and $t_0=L_0 / U_0=1593 \mathrm{~s}$ respectively, we set the magnetic diffusivity $\eta$ to $10^{12} {cm}^2/s$ . The simulation is integrated on a non-uniform grid with a strong refinement of $\Delta r=10^{-4}$ in the $0.1R_{\odot}$ region in code units and a coarser grid extending up to $20R_{\odot}$. The range for $\theta$ is $[\pi/2-0.145,\pi/2+0.305]$ with a stretched grid cell size of $\Delta \theta=2 \times 10^{-4}$, the total grid size is $1536 \times 1536$. We consider  reflective boundaries for the velocity components across $\theta_{min}$ and $\theta_{max}$. At $r=20R_{\odot} > r_{A,f}$ the fast magnetosonic point, we use an outflow boundary condition.

\begin{figure*}
    \centering
    
    \includegraphics[width=0.7\textwidth]{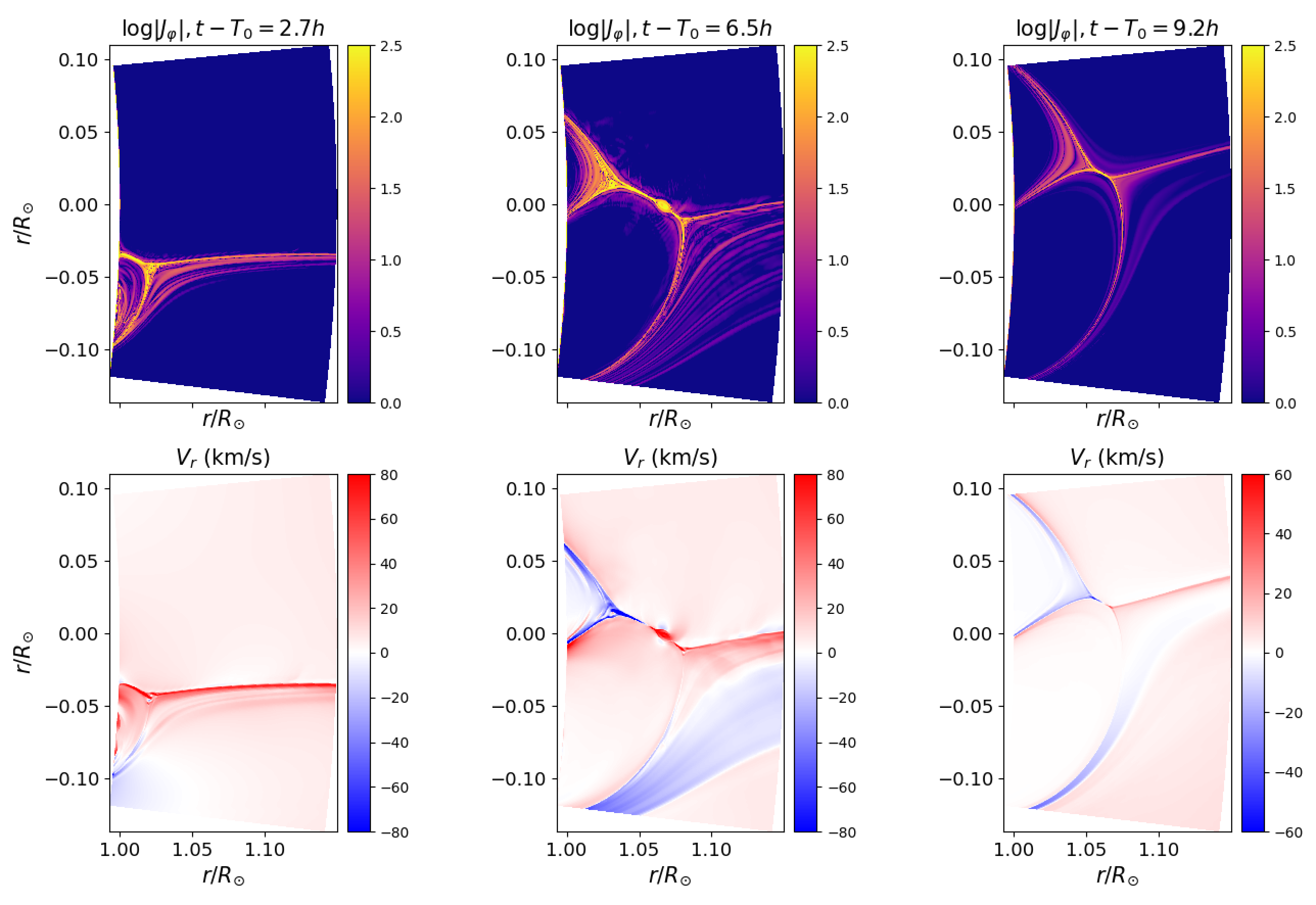}
    
    \caption{Development and decay of the current sheet during the flux emergence. The plot from left to right shows the logarithm of the out of plane current (top panels) and the radial velocity (bottom panels) at three different times. The first phase is a lengthening and thinning current sheet. The middle panel shows a well-developed CS undergoing reconnection through tearing. Finally, after the emergence, the CS starts decaying and reconnection stops. Velocity jets can be seen inward to the bipole footpoints and outward in the plume's fan.
    }
    \label{fig:emergence}
\end{figure*}

We initialize the atmosphere with a supersonic Parker wind and a purely radial field of 2G at the surface. We assume coronal conditions and do not include a chromosphere or transition region. We let the system reach a steady state before starting the flux emergence. For the inner boundary condition at $r=1 R_{\odot}$, we fix the temperature and density at the initial value $n=2 \times 10^8$ cm$^{-3}$ and  $T=1$ MK. The magnetic field is set to its background or emerging (see next) values, while the velocity field is computed to set the parallel electric field to zero. These boundary conditions are very close to the one described in \citep{Parenti2022}.

We then apply a time-dependent boundary condition for the magnetic field component ($B_r$,$B_{\theta}$) to control the emergence rate of the two polarities. A bipole of 20G amplitude and $\sim 10^\circ$ of latitudinal extent is projected on the spherical harmonics base (with $l_{\mathrm{max}} = 65$), and each coefficient is added to the background field in the boundary and increased linearly during the emergence phase. The spherical harmonics decomposition ensures that $\nabla \cdot B=0$ at all times. We focus on four setups with emergence rates of 3.61G/h, 4.52G/h, 6G/h and 9G/h, which are described in Figure \ref{fig:figsetup}. The duration of emergence for each setup is 5.53, 4.4, 3.3, and 2.2 hours, respectively.

In 2.5D, the Lundquist number $S=L v_A/\eta$ should be higher than a critical value of approximately $10^4$ to trigger the tearing instability \citep{Biskamp1986, 2007PhPl...14j0703L, 2009PhPl...16k2102B}. Hence, we set an explicit value of $\eta$ to be comfortably above this threshold (note that this depends on the length of the current sheet, and will vary with time as the flux emergence proceeds). But we must also be careful that the numerical resistivity remains smaller than or equal to the value of $\eta$ chosen. Based on our experience of the onset of the tearing mode with the PLUTO code \citep{Reville2020ApJL}, we find that a value of $\eta = 10^{12}$ cm$^2$/s and the above described grid resolution satisfy these requirements with $S_{\eta} \sim S_{\mathrm{num}} \sim 10^4$. We have also performed some simulations with $\eta = 10^{11}$ cm$^2$/s and $\eta = 10^{13}$ cm$^2$/s, and did not notice significant changes during the bursty reconnection phase (see Figure \ref{fig:periodi}).

\section{Results}\label{sec:Res}

\subsection{Study of the current sheet formation and aspect ratio} 

Figure \ref{fig:emergence} shows the time evolution of the simulation for a rate of emergence $3.61$G/h. We chose three characteristic phases of the emergence and relaxation phase, showing the logarithm of the out-of-the-plane current density and the radial velocity (top and bottom panels). In the early stages of emergence, the current sheet is created immediately and the tearing instability is triggered. The current sheet (CS) is disrupted several times, but continues to lengthen as the emergence continues and reaches its plateau. Close to the peak of the emergence (middle panel), reconnection occurs and plasmoids are ejected on both sides of the CS. Finally, after the emergence has stopped, the CS slowly decays, shortening, thus stopping the reconnection process. This stage can be seen in the right panel of Figure \ref{fig:emergence}.

\begin{figure}[ht]
    \centering
    \includegraphics[width=0.50\textwidth]{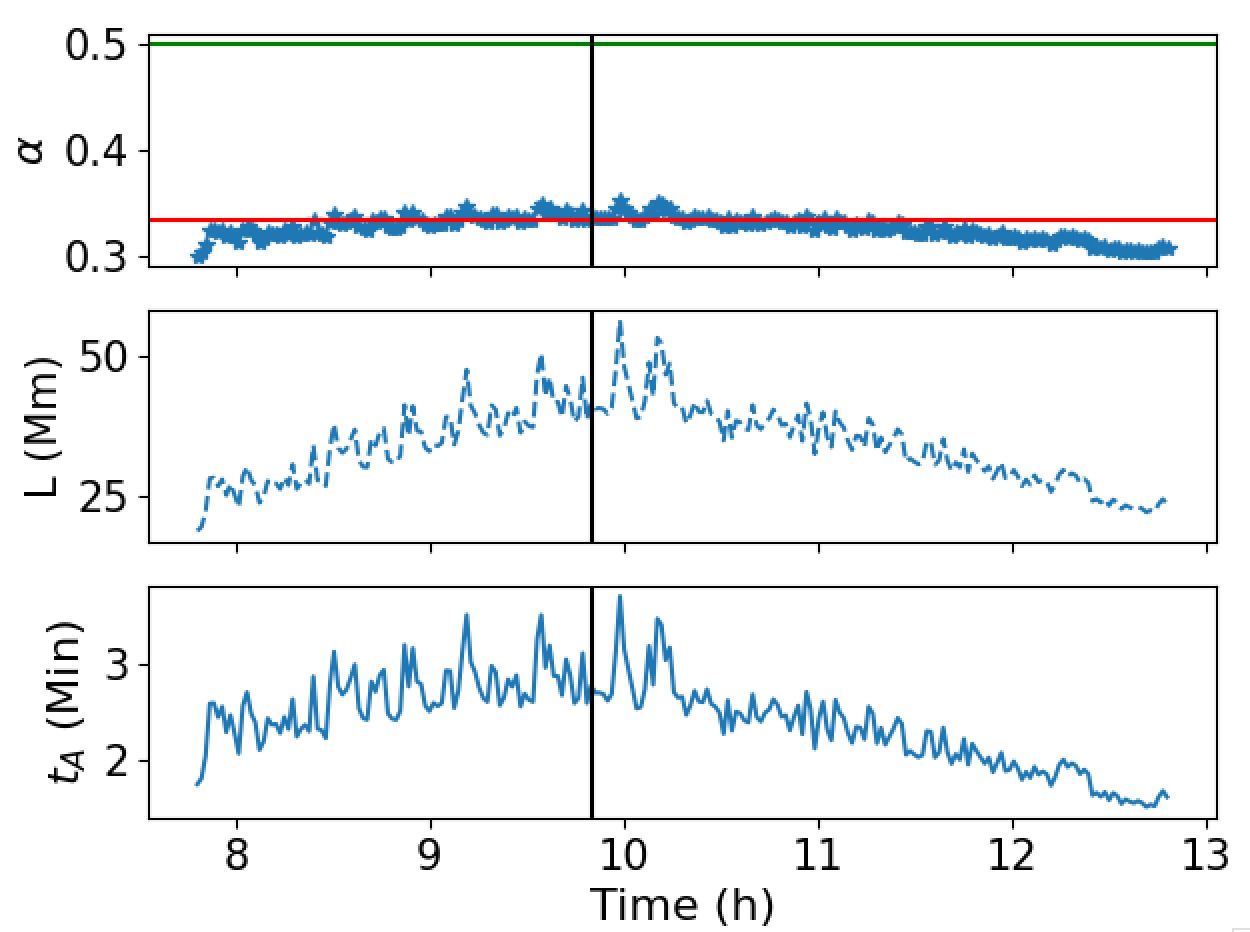}
    \caption{Evolution of the power law index $\alpha$,  L, and the Alfvén time $t_A$ during the flux emergence and relaxation phase with a rate of 4.52 G/h. The black vertical line indicates the time when the emergence is finished. The green horizontal line represents the critical $\alpha$ triggering the Sweet-Parker reconnection ($\frac{1}{2}$), and the red horizontal line denotes the critical $\alpha$ for the establishment of ideal tearing ($\frac{1}{3}$). The third panel shows the evolution of $t_A = L/v_A$. One can see that it remains close to 3 min during the whole emergence phase, while the current sheet length double in size and follow linearly the evolution of the Alfvén speed increase in the bipole.}
    \label{fig:tearing_10}
\end{figure}

The CS's length (L) is automatically measured by calculating $L=\mid\mathbf{B}\mid/ \mid\nabla\times\mathbf{B}\mid$ then fixing a threshold to locate the current sheet \citep[see, e.g.,][]{2022ApJ...935L..21N}. Figure \ref{fig:tearing_10} shows the evolution of the current sheet length obtained by this method.

\citet{2014ApJ...780L..19P} introduced the concept of "Ideal" tearing, in which the growth rate of the instability is independent of the Lundquist number. Assume that the ratio of the thickness of the current sheet $a$ to its length scale $L$ scales as $S^{-\alpha}$, S is the Lundquist number and $\alpha$ is the power law index. There is a critical value of $\alpha$ at which the growth rate is constant, $\gamma t_A \sim S^{\frac{-1+3 \alpha}{2}}=cnst$, which is equal to $1/3$. If $\alpha$ is greater than $1/3$, the growth rate tends to diverge with increasing Lundquist number, while if $\alpha$ is less than $1/3$, the growth rate tends to zero. We thus expect the reconnection to occur precisely at $\alpha=1/3$, when the current sheet forms from large aspect ratios.

In the first panel of Figure \ref{fig:tearing_10}, we show the estimated value of $\alpha$ as a function of time. We notice that during the flux emergence phase, $\alpha$ increases and reaches values very close to $1/3$,  when reconnection begins. This suggests that the tearing mode is effectively ideal, and that the reconnection rate obtained in the simulations should be close to realistic values. Interestingly, as shown in Figure \ref{fig:tearing_10}, the Alfvén time $t_A = L/v_A$, varies much less than the current sheet length, and remains close to 3 minutes throughout the emergence phase. This is due to the linear dependence of $L$, that follows the linear increase of $v_A$ during the emergence ($v_A$ is computed in the bipole, away from the current sheet, as in usual tearing mode analysis). The value of $t_A$ is also constant for different emergence rate, which will have consequences on the measured periodicity of the reconnection jets (see section \ref{sec:floats}).

Once the flux emergence phase is complete, we observe a decrease in both the current sheet length and the value of $\alpha$. This indicates that the magnetic field is settling and converging to the X-point. The decrease in $\alpha$ indicates that the CS (current sheet) becomes thicker and more diffuse, and the magnetic field lines are less tightly packed. Once the current sheet starts thickening, the tearing reconnection essentially stops.

\begin{figure*}
    \centering
    \includegraphics[width=0.90\textwidth]{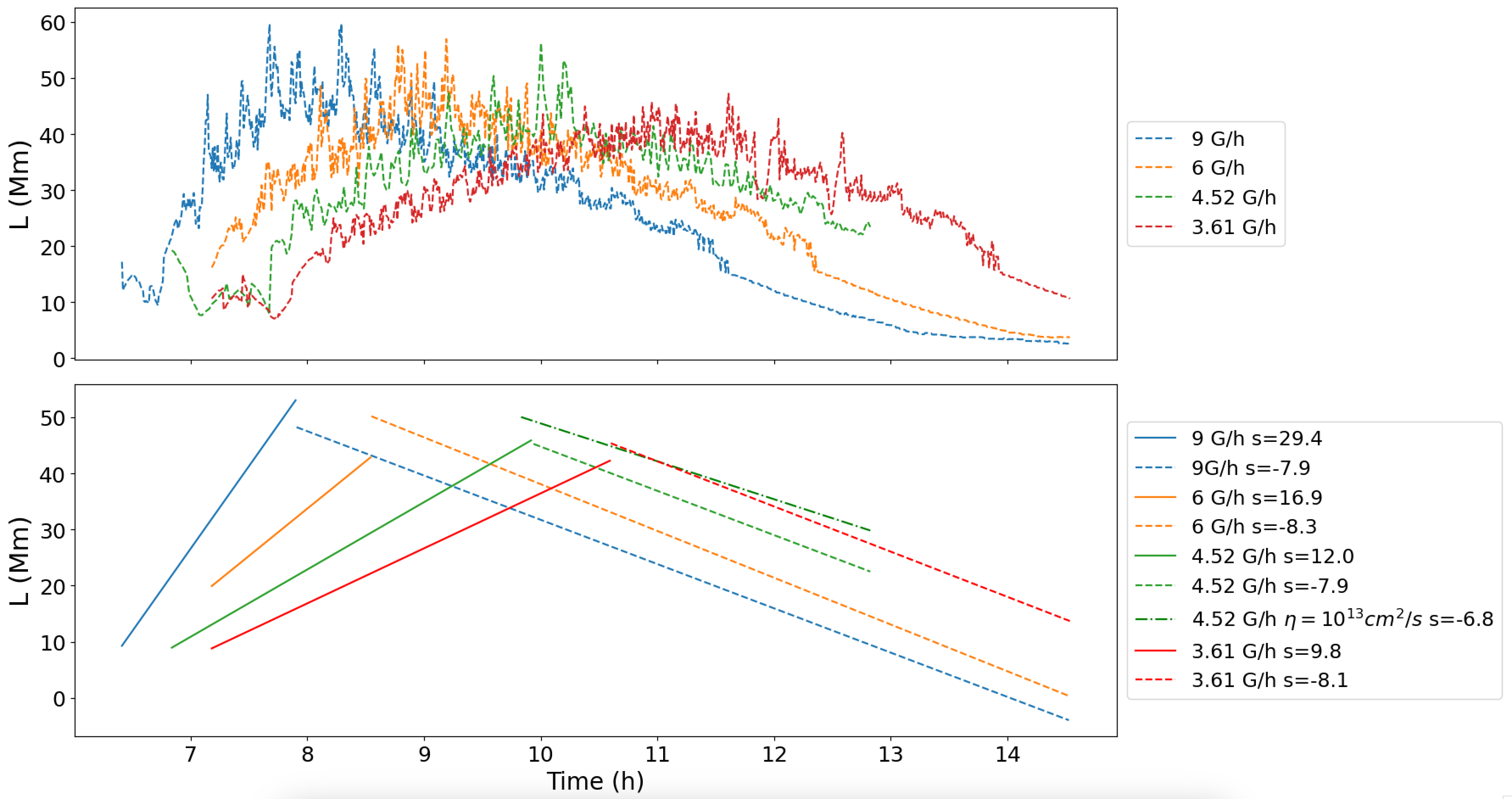}
    \caption{In the upper panel, we plot the evolution of current sheet length (L) for all setups with $\eta=10^{12} cm^2/s$ and the second panel shows the linear fit with the corresponding slope "s" .}
    \label{fig:L_T}
\end{figure*}

As the emergence phase ceases, a quasi-steady reconnecting phase begins, during which the CS starts to diffuse. To quantify the behavior of the current sheet, we calculated L over time for all our simulations, as illustrated in Figure \ref{fig:L_T}. There is a clear linear relationship between L and the rate of flux emergence, suggesting that the rate of emergence is a key factor in the evolution of L. Furthermore, we observed that the decay of the CS followed a linear trend with the same slope for the same magnetic diffusivity. As expected for quasi-steady reconnection, the decay rate depends on the chosen $\eta$ value: a higher/lower $\eta$ leads to a slower/faster decay of L \citep{2022LRSP...19....1P}.

\subsection{Formation and propagation of jets and reconnection plasmoids}\label{sec:floats}

\begin{figure}[h]
    \centering
    \includegraphics[width=0.50\textwidth]{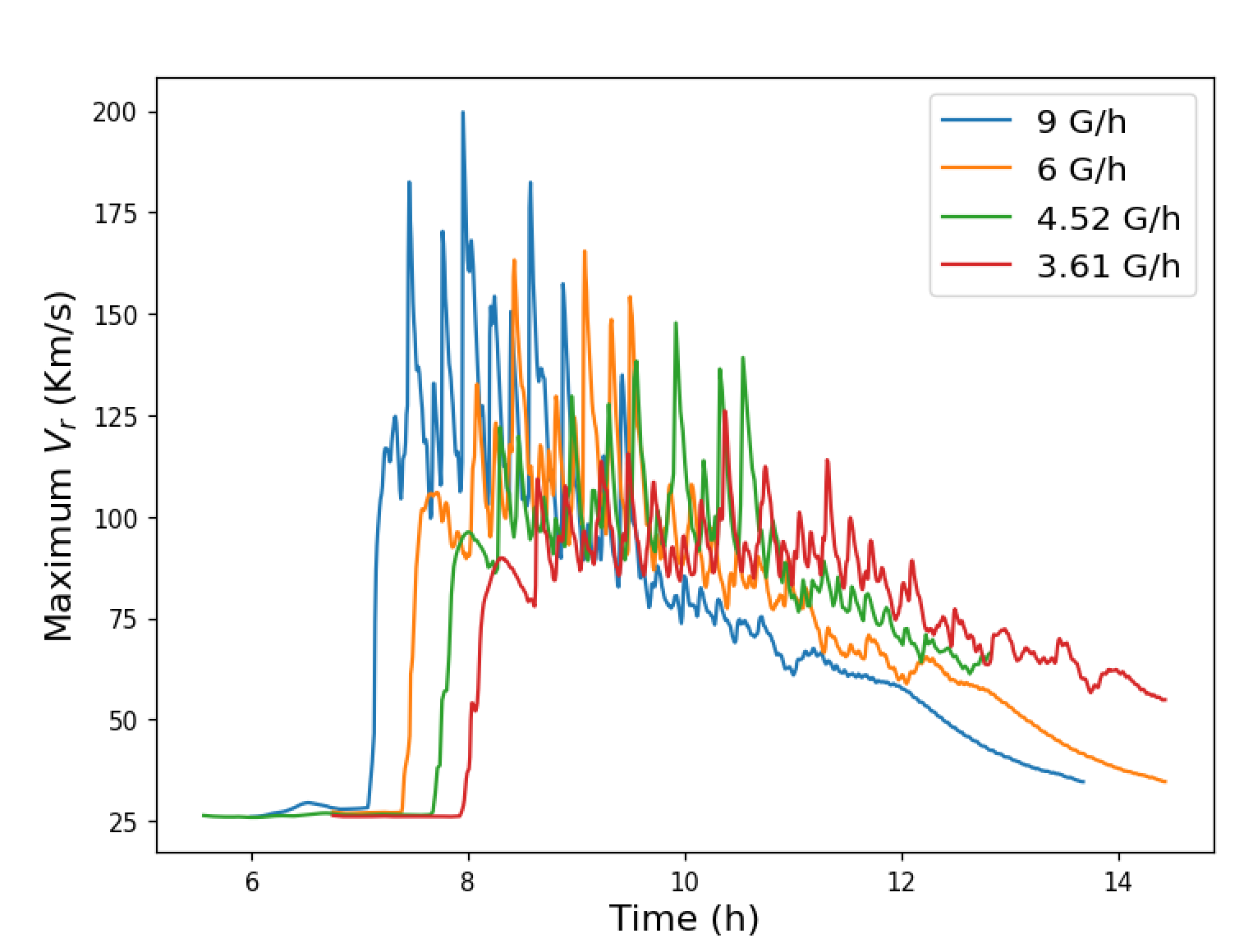}
    \caption{The maximum values of the radial velocity simulated in the plume above the current sheet at a distance of approximately $1.5 R_{\odot}$ for the different setups of flux emergence. Magnetic diffusivity was fixed in all simulations at $\eta=10^{12} cm^2/s$.}
    \label{fig:maxprofiletot}
\end{figure}

In all the setups, we observed recurrent jets and velocity spikes. Plasmoids repeatedly form and are ejected from the current sheet, triggering sequences of perturbations of the plasma feeding the solar wind above the cusp of the forming pseudo-streamer. 
In Figure \ref{fig:maxprofiletot}, we present the maximum value of the radial velocity for each setup of flux emergence rate. It shows that flux emergence rate has a direct impact on the speed of the triggered jet, which varies between  $50$ and $200$ km/s. The higher the emergence rate the higher the amplitude of the jet. Furthermore, the reconnection process heats the plasma and creates density structures that can be related to EUV observations. 
\subsubsection{Synthetic EUV emission}
\begin{figure}[!htpb]
    \centering
    \begin{minipage}[b]{0.6\linewidth}
        \centering
        \includegraphics[width=\linewidth]{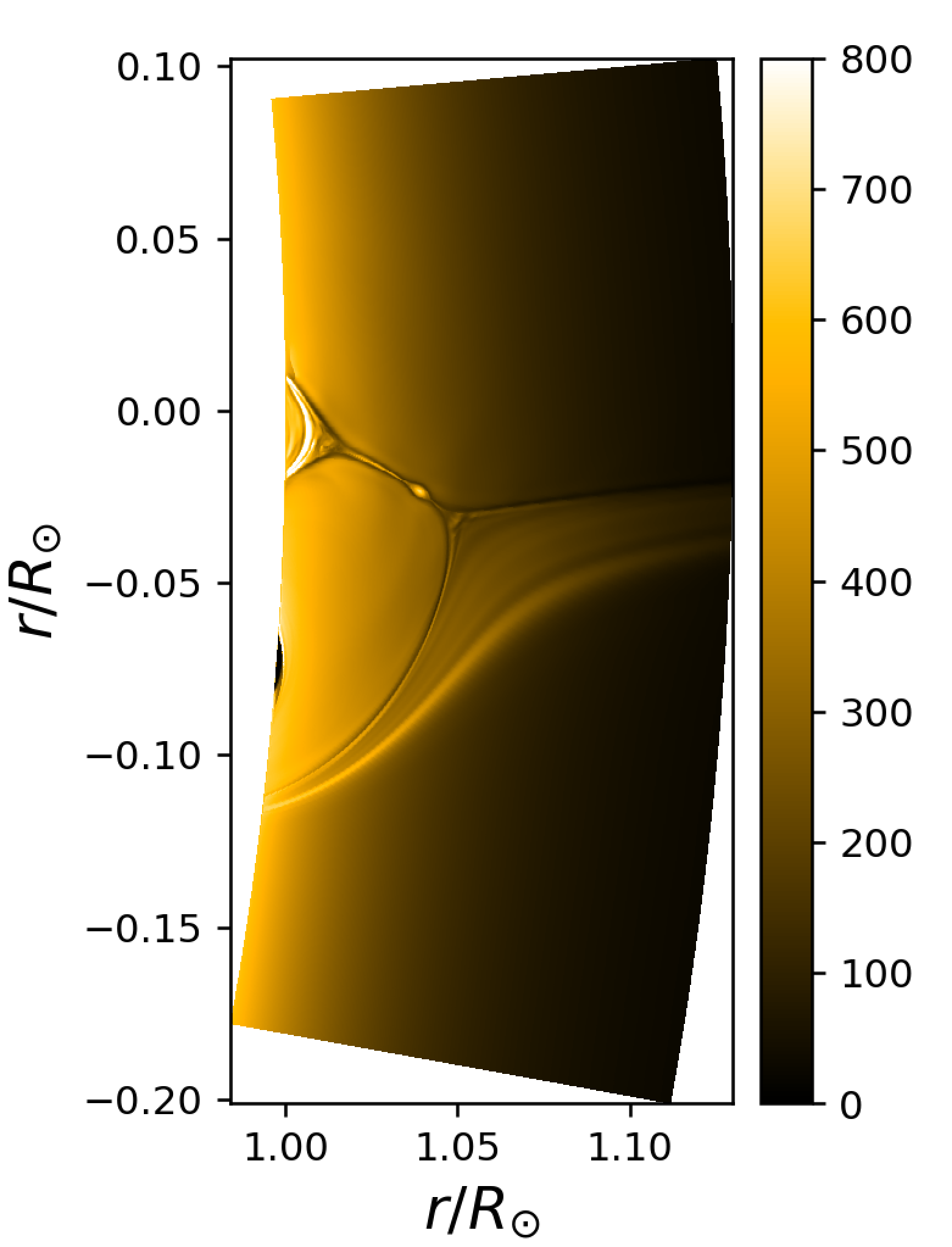}
        \par\medskip
        (a) $\SI{171}{\angstrom}$
    \end{minipage}
    \vfill
    \begin{minipage}[b]{0.63
\linewidth}
        \centering
        \includegraphics[width=\linewidth]{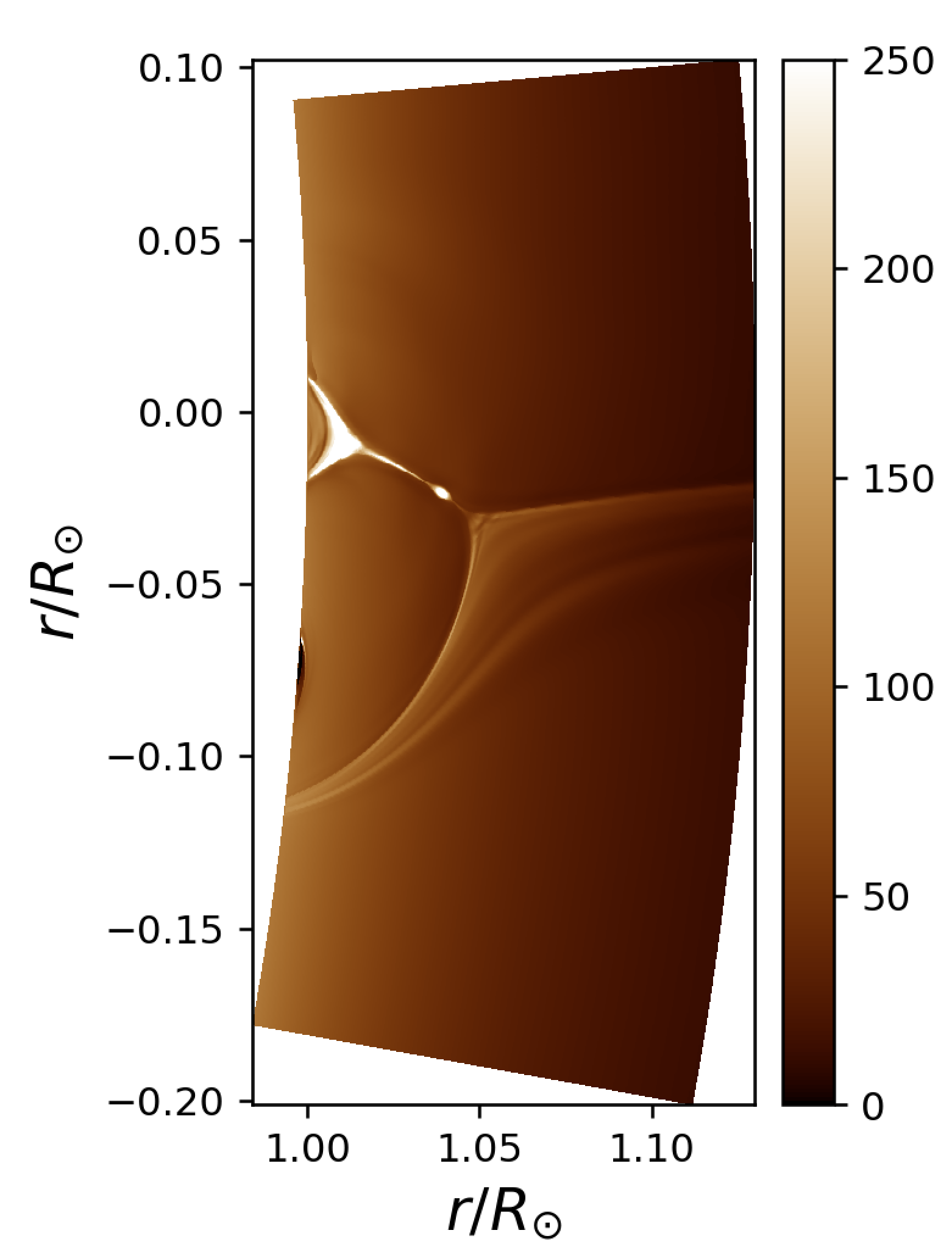}
        \par\medskip
        (b) $\SI{193}{\angstrom}$
    \end{minipage}
    \caption{Synthetic EUV emission intensity maps produced from the simulation results for two wavelengths observed by the Solar Dynamics Observatory. These emission intensity maps were produced from the simulation setup with a flux emergence rate of $4.52 G/h$.}
    \label{fig:Emexample}
\end{figure}

A $2.5D$ MHD simulation provides electron densities and temperatures $(r,\theta, \phi)$ inside a 2D plane. We can construct a 3D cube necessary to compute EUV images by multiplying the 2D plane over a depth $L_{los}$ along the line of sight assumed perpendicular to the 2D simulation plane. The emission $(dI_j)$ from the plasma in each cell (j) of the cube can then be calculated. For the Solar Dynamic Observatory (SDO) EUV bands, the dominant process assumed for the observed emission is excitation by electron-ion collisions followed by spontaneous emission. The expression for the emission is given by:
\begin{equation}
    dI_j = A_f \cdot G(T_j, n_j) \cdot n_j^2 \cdot dV
\end{equation}

Here, $A_f$ represents the spectral response function of the instrument being simulated, $G(T_j, n_j)$ encapsulates the atomic physics involved in the spectral line formation and is dependent on the local electron density (n) and temperature (T). The values of $A_f$ are provided by the Solar Dynamics Observatory/Atmospheric Imaging Assembly instrument team (SDO/AIA) and distributed in the SolarSoft library, while $G(T_j, n_j)$ is the contribution function calculated using the CHIANTI atomic physics database version 7.1.3 and ChiantiPy interface \citep{1997A&AS..125..149D,2013ascl.soft08017D} by assuming ionization equilibrium and coronal composition. The values of $n_j$ and $T_j$ are derived from the output of the MHD model.

The simulated AIA images are generated in units of $Dn/s$, which is the calibrated data unit. Using this method, the synthetic images should capture the emission properties of the corona and allow comparison with the observational data obtained by SDO/AIA. Figure \ref{fig:Emexample} shows the EUV emissions observed in $\SI{193}{\angstrom}$ and $\SI{171}{\angstrom}$. We can see clearly that the  plasmoid is discernible in $\SI{193}{\angstrom}$. For this reason, the $\SI{193}{\angstrom}$ band was chosen due to its response function peaking at a temperature of approximately 1.5 MK reached by the plasmoids.

\begin{figure}[ht]
    \centering
    \includegraphics[width=0.50\textwidth]{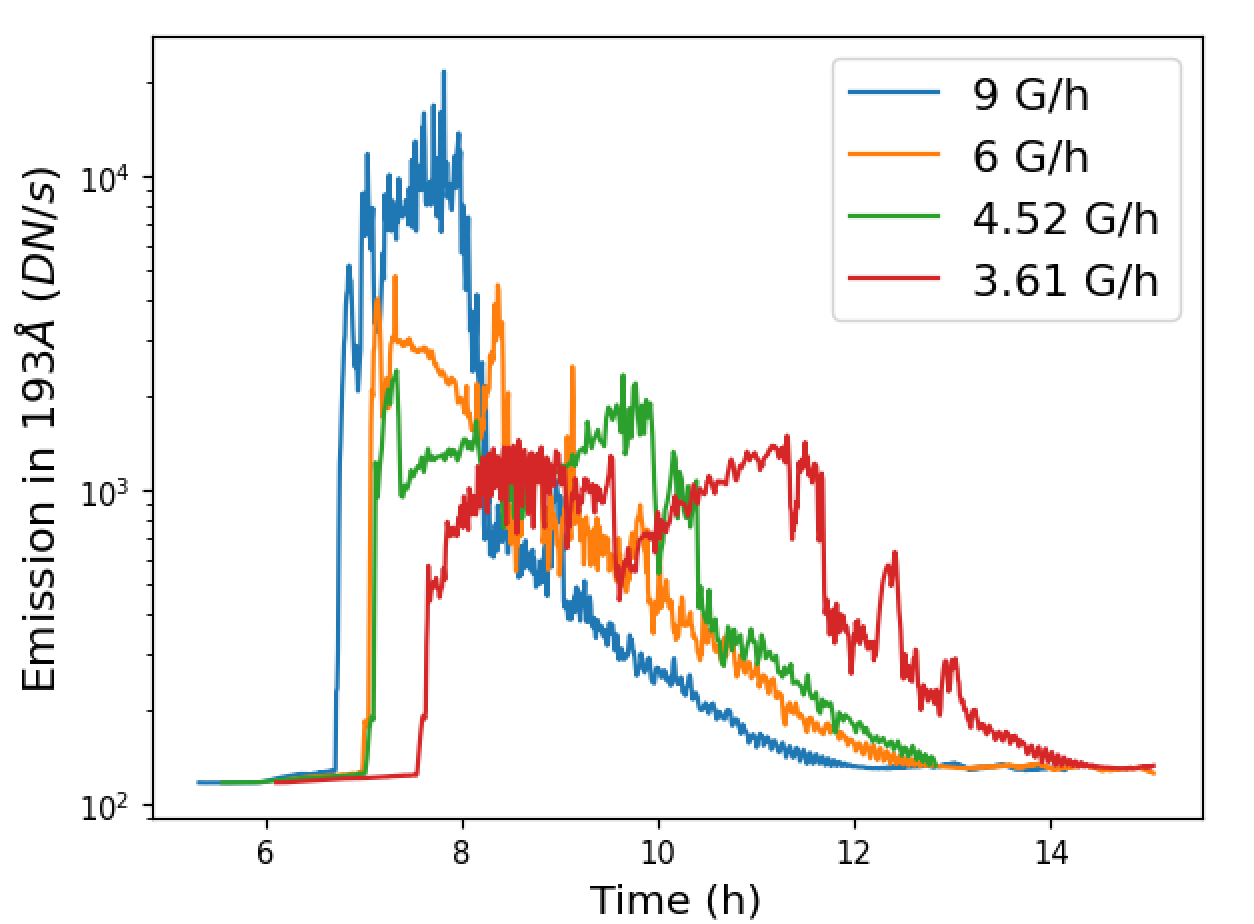}
    \caption{Synthetic emission intensities in the $\SI{193}{\angstrom}$   EUV line for the four emergence rates simulated for a fixed $\eta=10^{12} cm^2/s$}
    \label{fig:emissionEm}
\end{figure}

Figure \ref{fig:emissionEm} show the integrated intensity of the emission in $\SI{193}{\angstrom}$ for, respectively, several values of magnetic diffusivity and different flux emergence rates where we have captured the plasmoids more clearly so that we can focus on the emission triggered by magnetic reconnection. 

We observe that the rate of flux emergence influences the amplitude of brightening in EUV $\SI{193}{\angstrom}$. A fast magnetic flux emergence rate of 9 $G/h$ results in an emission amplitude of 3.$10^4$ Dn/s, while a slow flux emergence rate of $3.61 G/h$ results in an EUV brightness that is ten times smaller.

\subsubsection{Periodicity of jetlet-associated brightenings versus periodicity of outflows}

Our simulations reveal that the emerging bipole exhibits rapid brightening in the EUV $\SI{193}{\angstrom}$   channel as it interacts with the overlying magnetic field. This brightening is mainly driven by magnetic energy being converted into heat in the current sheet layer (ohming heating), as well as increased densities in the sequence of outflowing jets via magnetic reconnection. We have checked whether the brightening could arise from adiabatic compression rather than reconnection associated heating. Within the current sheet, the component $-p\nabla \cdot \mathbf{v}$ predominantly cools the local plasma, implying that the major contribution to temperature rise is indeed due to the ohmic heating term $\eta J^2$ during magnetic reconnection. EUV brightenings and jets should consequently be considered as the macroscopic signatures of local and bursty energy releases that develop along the reconnecting layer at the surface of the emerging bipole. \\

\begin{figure}
    \centering 
    \includegraphics[width=0.48\textwidth]{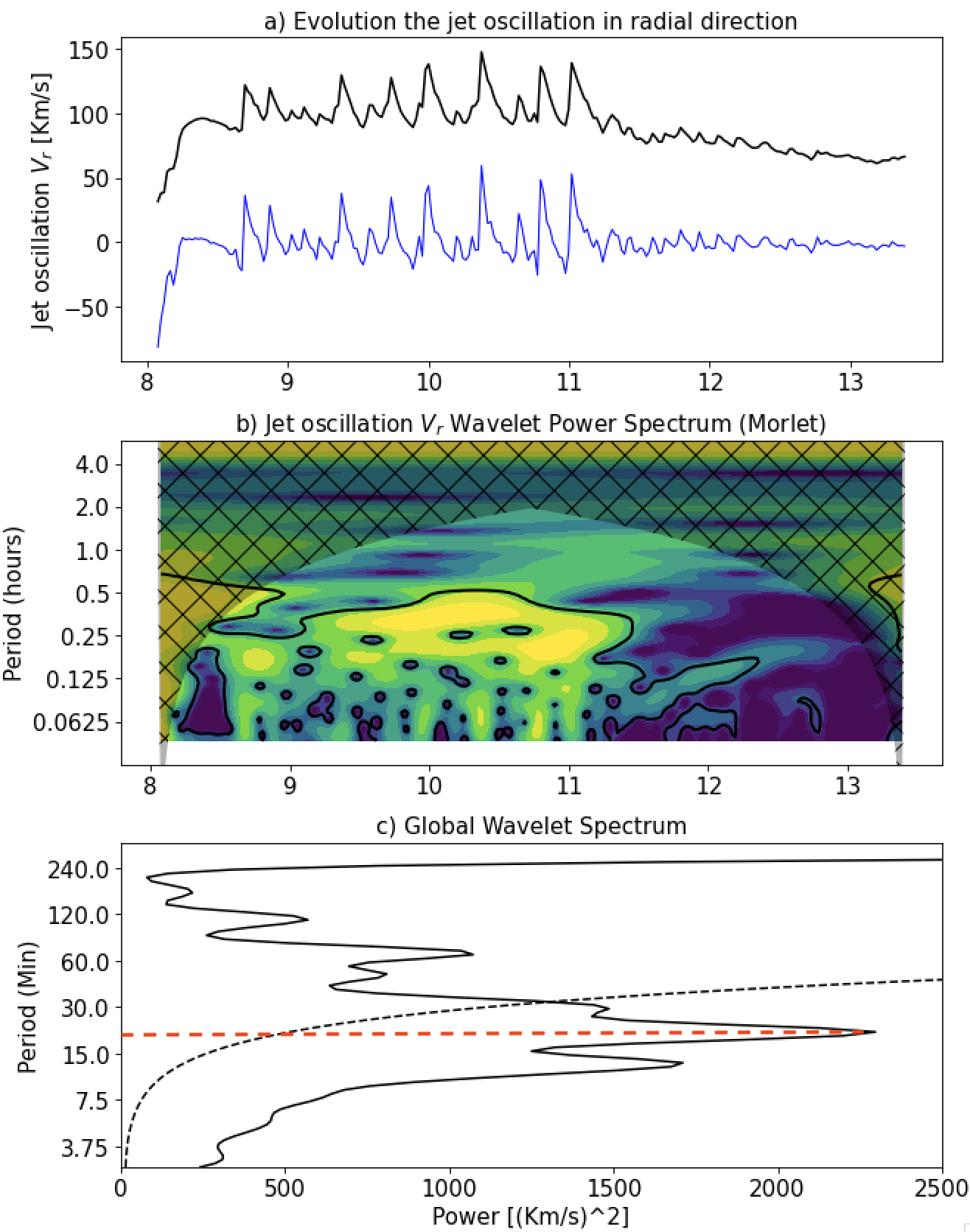}
    \caption{Wavelet analysis of the radial velocity for the setup of flux emergence 4.52 G/h and $\eta=10^{12} cm^2/s$ (a) Radial velocity evolution over time in $1.5R_{\odot}$,the detrended radial velocity after subtracting it from the original gray (color). (b)The wavelet power
    spectrum of the detrended signal is displayed and (c) the global wavelet power spectrum }
    \label{fig:Wavelet}
\end{figure}

Investigating the quasi-periodicity of these energy releases is important because it provides insights into the relationship between energy releases and the rate of magnetic flux emergence, which drives the underlying footpoint exchange mechanism studied here. We performed a wavelet analysis using the wavelet software package of \citet{1998BAMS...79...61T} to study the periodicity of the oscillating emission intensity and the radial velocity $V_r$ of the jets for the different flux emergence rates. Figure \ref{fig:Wavelet} shows the wavelet spectral analysis of the radial velocity $V_r$ for a flux emergence with a rate of 4.52 G.h$^{-1}$. The signal is first detrended by subtracting a linear polynomial fit from the original data. The resulting detrended signal is then normalized by dividing it by its standard deviation to ensure that it has a zero mean and unit variance.

\begin{figure*}[!htpb]
    \centering
    \begin{minipage}[b]{0.45\textwidth}
        \centering
        \includegraphics[width=\textwidth]{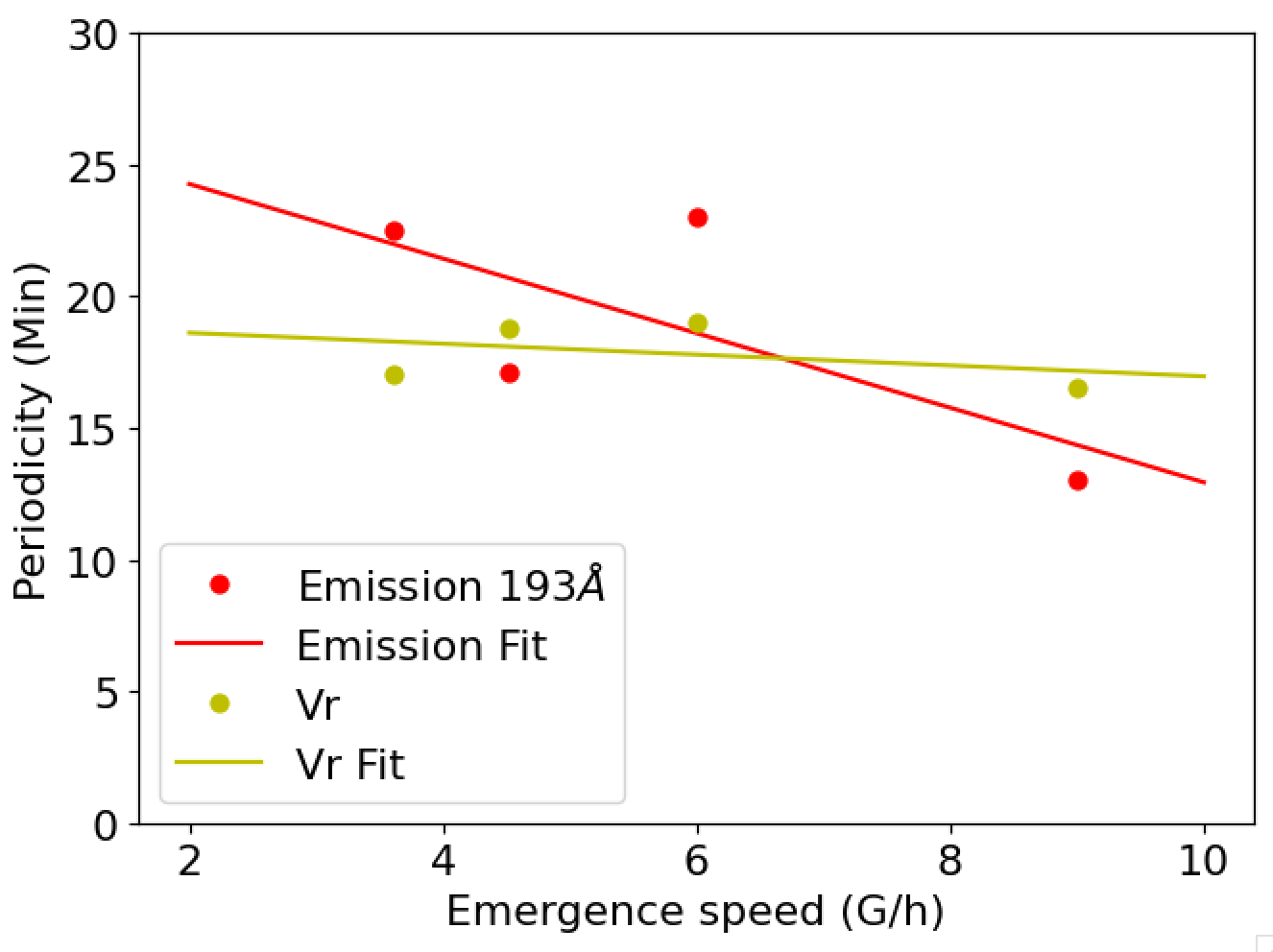}
        \par\medskip
        (a)
    \end{minipage}\hfill
    \begin{minipage}[b]{0.45\textwidth}
        \centering
        \includegraphics[width=\textwidth]{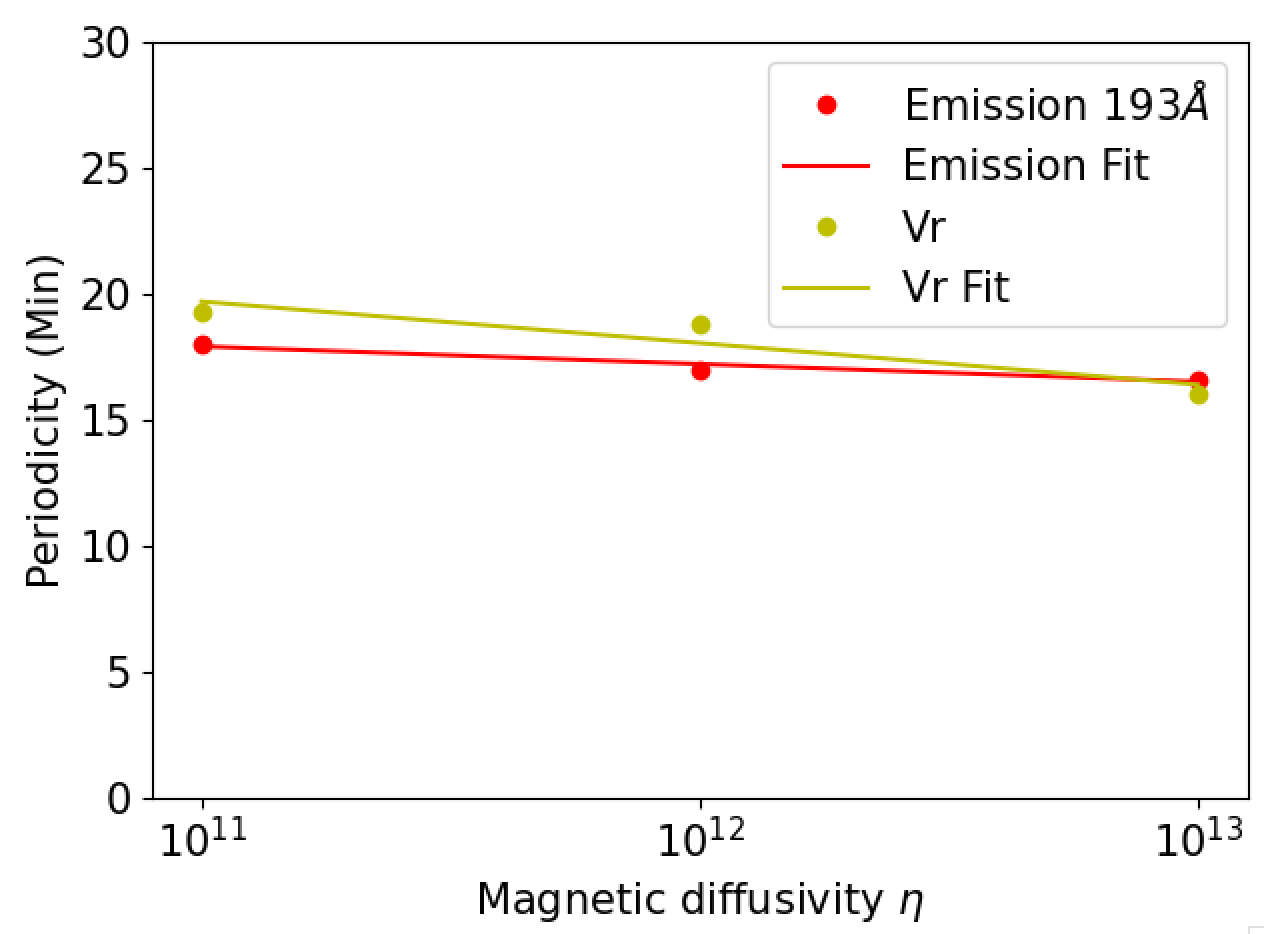}
        \par\medskip
        (b)
    \end{minipage}
    \caption{Figure depicting the periodicity of oscillation of radial velocity and EUV $\SI{193}{\angstrom}$ for (a) different emergence rates (G/h) but a fixed diffusivity of $10^{12} \, \text{cm}^2/\text{s}$ (b) different magnetic diffusivity values for the same emergence rate $4.52 \, \text{G/h}$.}
    \label{fig:periodi}
\end{figure*}

Figure \ref{fig:periodi} shows the periodicity of radial velocity spikes and the $\SI{193}{\angstrom}$ EUV emission for several flux emergence rates and for several magnetic diffusivity values. First, we notice that the radial velocity oscillation is mostly unchanged with the flux emergence rate. Second, we observe a good correspondence between the periodicity of the jet outflows and the EUV brightenings. This is expected because, in the simulations, the two processes are directly related to the tearing induced reconnection phase. It has been, however, very surprising that the periodicity does not change significantly with the control parameters of the simulation.

This relatively constant periodicity across the various simulations parameters can be interpreted as follows: the Alfvén time $t_A$ away from the current sheet is approximately 3 minutes, and as we are in the ideal tearing reconnection regime $\gamma t_A$, the normalized growth rate, is close to unity. Hence, once the CS has reached the ideal tearing ratio, the periodicity of jets is directly related to the time for the CS to be disrupted, i.e., a few times $t_A$, which leads to the observed 19 minutes. Moreover, as shown in Figure \ref{fig:tearing_10}, the Alfvén time stays roughly constant throughout the emergence phase, as the current sheet lengthen and adapt to the increasing magnetic field strength in the bipole. Indeed, in our setup it is strictly equivalent to increase either the emergence speed or the bipole amplitude, and thus $t_A$ remains close to this 3 minute value in all the setups. Note that as $S \sim 10^4$, the $\gamma t_A \sim cnst$, may not be fully reached yet. 

Finally, plasmoids propagate within the CS either sunward and anti-sunward, while only anti-sunward plasmoid will launch jets propagating in the solar wind. As a result, the periodicity of the velocity spikes may be higher than the full current sheet disruption and reformation cycle, as the jets correspond only to the outward-moving plasmoids. This nonetheless suggests that the periodicity of EUV brightenings and velocity jets may follow a universal rule and depend only on the local value of $t_A$.

These results are also consistent with previous observations reported by \citet{2021ApJ...907....1U} and \citet{2023ApJ...951L..15K}, who also found that the oscillations in the plume exhibit a range of periods similar to those of the jet (10 to 20 minutes).

\newpage

\subsection{Alfvén waves versus magnetoacoustic waves: Which wave mode dominates the coronal jets?}
There has been a debate in the community whether propagation disturbances (PDs) in plumes observed in the solar corona are plasma outflows or slow-mode waves \citep{2015LRSP...12....7P,2016GMS...216..395W}.
The work of \citet{2016GMS...216..395W} suggests that both of these interpretations may be correct, as reconnection at the footpoints of the plumelet drives flows that can be observed as "jetlets" and generate Alfvénic fronts, but the dense material in the jets travels more slowly and an inhomogeneous wake of shear and compressible turbulence should be observed between the jet and the Alfvénic front. \\

In our simulations, the flux emergence increases L, and when the tearing instability is triggered, plasmoids are then generated that move up and down the X-line. It is important to understand what type of wave this process would generates, transverse Alfvén wave or magnetoacoustic waves? To check this, we start with the linearized equation of ideal magnetohydrodynamics (MHD). Three well-known modes can be distinguished: fast and slow magnetoacoustic waves, as well as (transverse) shear Alfvén waves.\\
For shear Alfvén wave, we have:

\begin{equation}
\begin{cases}
v_{\mathrm{pA}}=\frac{\omega}{k}=v_{\mathrm{A}} \cos \theta \cong v_{\mathrm{A}}\\
\frac{\delta \boldsymbol{v}}{v_{\mathrm{A}}}= \pm \frac{\delta \boldsymbol{B}}{B_0} \\
\delta \rho=0\\
\delta \mid \mathbf{B} \mid =0 
\end{cases}
\end{equation}

Where $v_{\mathrm{A}}$ is the Alfvén speed, $\theta$ is the angle between the wavevector $\boldsymbol{k}$ and the magnetic field $\boldsymbol{B}$ and is close to zero, $\delta \boldsymbol{v}$, $\delta \boldsymbol{B}$, and $\delta \rho$ are the perturbed plasma velocity, magnetic fields, and plasma density, respectively, and $B_0$ is the background magnetic magnitude.
For slow and fast waves, the phase speeds are

\begin{equation}
\begin{aligned}
v_{\mathrm{ph} \pm}^2 & =\left(\frac{\omega}{k}\right)^2 \\
& =\frac{1}{2}\left(v_{\mathrm{S}}^2+v_{\mathrm{A}}^2\right) \pm \frac{1}{2}\left[\left(v_{\mathrm{S}}^2+v_{\mathrm{A}}^2\right)^2-4 v_{\mathrm{S}}^2 v_{\mathrm{A}}^2 \cos ^2 \theta\right]^{1 / 2}
\end{aligned}
\end{equation}

To simplify our analysis, we focus on the case where $v_{\mathrm{S}} << v_{\mathrm{A}}$ which verified when $r<R_A$ (below the Alfvén radius). We then have for fast magnetoacoustic waves \citep{1975RvGSP..13..263H} :
\begin{equation}
\begin{cases}
 \frac{\delta \rho}{\rho_0} =\frac{\delta \mid  \mathbf{B} \mid }{B_0}\\

v_{ph+}{ }^2 \cong v_A^2 \\
 \mid \frac{\delta \rho}{\rho_0} \mid< 0.1 \\

\end{cases}
\end{equation}
and for slow magnetoacoustic waves:
\begin{equation}
\begin{cases}
\mid \frac{\delta \rho}{\rho_0} \mid=\mid \frac{\delta \mathbf{v}}{v_S}\mid\\
v_{ph-}{ }^2 \cong v_s{ }^2 \cos ^2 \theta\\
 (\mid \frac{\delta \rho}{\rho_0} \mid> 0.1)  \\
\delta \mid \mathbf{B} \mid  \cong 0
\end{cases}
\end{equation}

\begin{figure}[ht]
    \centering
    \includegraphics[width=0.45\textwidth]{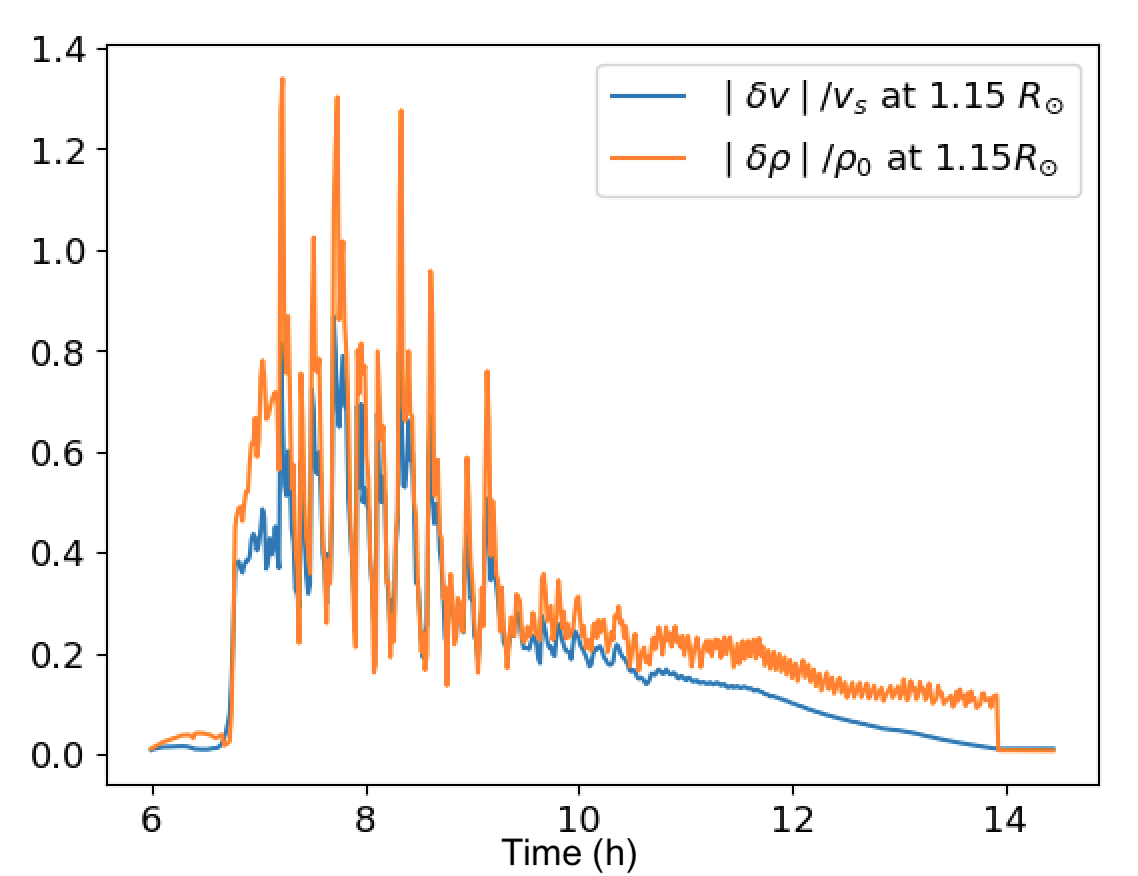}
    \caption{Density and velocity perturbations as a function of time for the emergence rate 9 G/h at 1.15 $R_{\odot}$ in the jet's stalk.}
    \label{fig:slowmag}
\end{figure}

Figure \ref{fig:slowmag} presents the plasma velocity fluctuation and density fluctuation at 1.15 $R_{\odot}$.It is evident that $\mid \frac{\delta \rho}{\rho_0} \mid \sim \mid \frac{\delta \mathbf{v}}{v_S}\mid$. Given the additional constraint that $\mid \frac{\delta \rho}{\rho_0} \mid> 0.1$ and that we have checked that  $\mbox{Max}(\mid \delta \mid \mathbf{B} \mid \mid) \cong 10^{-5} \ll 1$, the characteristics exhibited by the jets align remarkably well with the behavior expected from slow magneto-acoustic waves. 

\begin{figure}[!h]
    \centering
    \rotatebox{90}{ $ \hspace{2.5cm} {r/R_{\odot}} $}
    \includegraphics[width=0.44\textwidth]{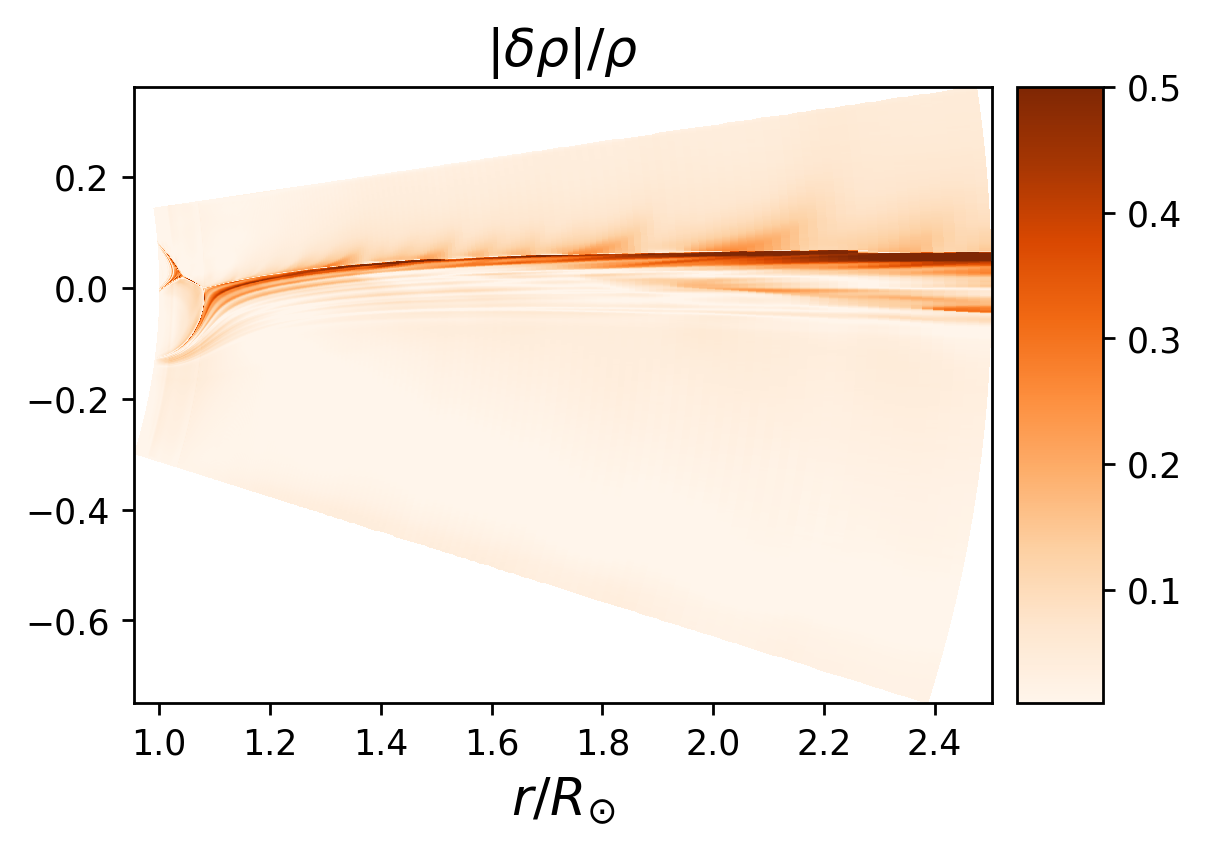}
    \caption{2-D map of density fluctuations ($\delta \rho$) relative to the local density $\rho$. $\delta \rho$ is computed by subtracting two consecutive frames 16 minutes time-lapse}
    \label{fig:RHO}
\end{figure}

We show in Figure \ref{fig:RHO} a 2-D map of the density variations relative to the local density in the high corona and the nascent solar wind. These density variations suggest compression waves are triggered and propagate along the plume that has formed above the cusp of the emerging bipoles. These density fluctuations originate in the plasmoids ejected outwards during the magnetic reconnection process. As the size of magnetic islands becomes larger, the medium surrounding the plasmoids is compressed by the outflowing structure, leading to changes in the density.

We now focus on the Alfvénic character of the generated waves. We define the cross helicity, $H_c$, as
\begin{equation}
H_c=\frac{2 \cdot (\delta \mathbf{v_A} \cdot \delta \mathbf{v})}{\delta \mathbf{v_A}^2+\delta \mathbf{v}^2}, 
\end{equation}

$H_c$ quantifies the degree of correlation in the fluctuating velocity and magnetic field components. This provides information on the nature and direction of the propagating waves. In the case of pure shear Alfvén waves, we have $\delta v=\pm \delta v_A$, where $\delta v_A $ is the perturbed Alfvén velocity, and $H_c =\pm 1$. The propagation direction of Alfvén waves is indicated by the sign of the cross helicity. A '+' sign denotes propagation antiparallel to the local mean magnetic field $B_0$, while a '-' sign denotes propagation parallel to $B_0$. In the solar wind, particularly in the inner heliosphere, Alfvén waves predominantly propagate outward from the Sun. As our background field is positive, negative cross helicity indicates fluctuations propagating outward into the heliosphere, while positive values denotes inward-propagating fluctuations. 

\begin{figure*}
    \centering
    \includegraphics[width=0.8\textwidth]{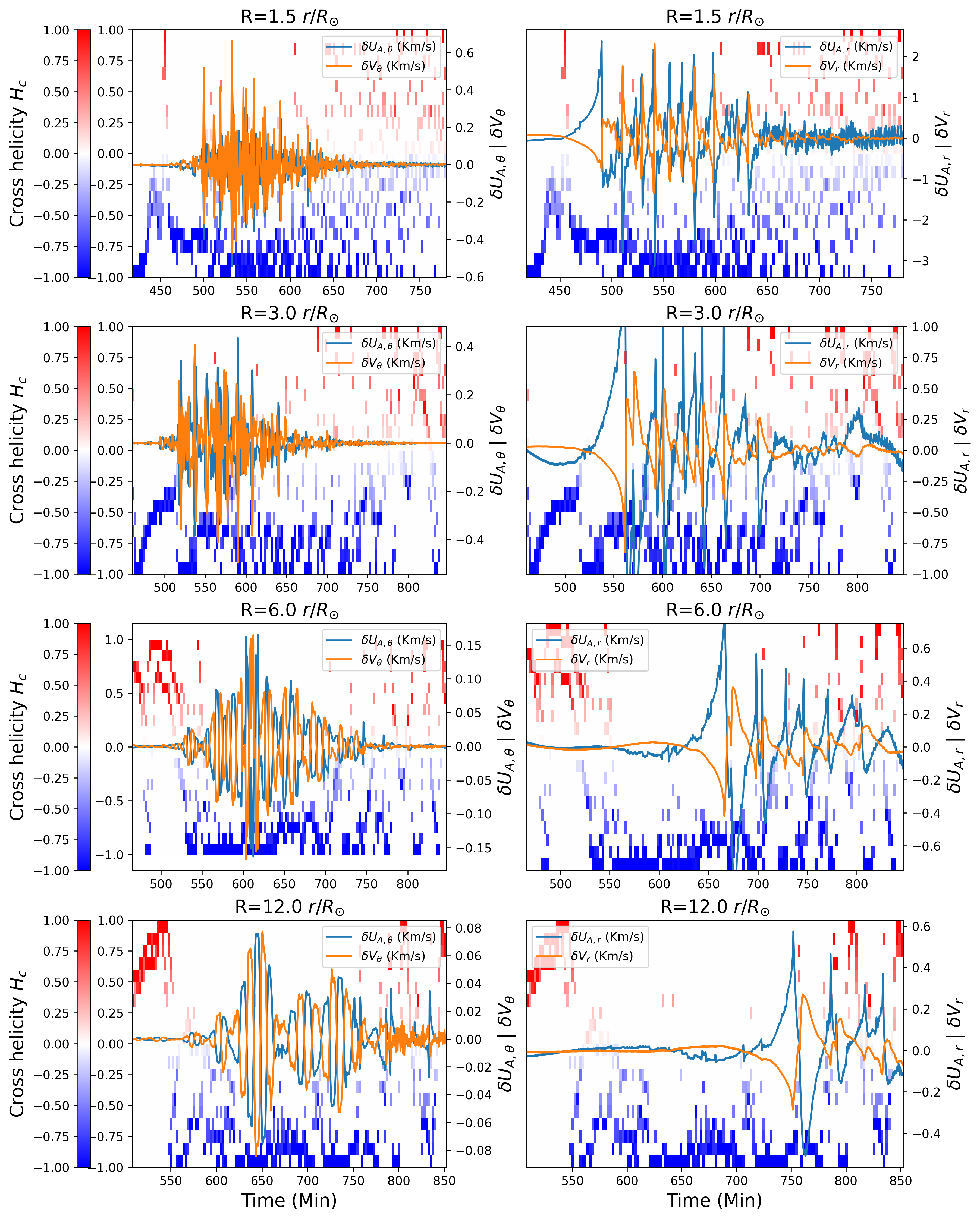}
    \caption{Radial and tangential Alfvénic and velocity perturbations at $r/R_{\odot}$=[$1.5$,  $3$, $6$, $12$] for the emergence rate of $9 G/h$. Radial Alfvénic perturbations (represented by the blue line) and radial velocity perturbations (represented by the orange line) at various radial locations in the solar corona. The plot also displays the cross-helicity (represented by the red and blue colormap) at these locations.}
    \label{fig:Sss}
\end{figure*}



We use the maximum radial velocity to pinpoint the jet's location. Subsequently, we calculate perturbations in magnetic field and plasma velocity, both in the radial and tangential directions. A linear function of time is employed to fit the background profile, which is then subtracted from the original data. The resulting perturbed signal is normalized by dividing it by its standard deviation, resulting in a zero mean and unit variance, denoted as $\delta Z$.

In Figure \ref{fig:Sss}, we present a time series illustrating radial and tangential perturbations in Alfvén and plasma velocities at various radial positions for a simulation with an emergence rate of 9G/h. We selected four heliocentric radial distances [1.5, 3, 6, 12] in R$_\odot$ to monitor the jet's evolution. The colormap represents cross helicity, with blue/red colors indicating negative cross helicity during the passage of the wave packet, signifying outward-moving Alfvén waves.

The tangential perturbation exhibits a decreasing trend with solar radius, transitioning from 0.6 km/s at 1.5R$\odot$ to 0.08 km/s at 12R$\odot$. A similar trend is observed in the radial direction, with velocity decreasing from 2 km/s to 0.6 km/s. This behavior can be attributed to the diffusive nature of slow magneto-acoustic waves and associated damping, as well as the damping due to decreasing resolution at higher altitudes.

As we ascend in altitude, the jet displays reduced compressibility. Significant disparities exist between tangential and radial perturbations. Tangential perturbations are characterized by a more pronounced anti-correlation as well as higher frequencies. Notably, when examining various altitudes 1.5$R_{\odot}$, 3$R_{\odot}$, 6$R_{\odot}$ and 12,$R_{\odot}$, we observe distinct patterns in the tangential perturbations.

At 1.5$R_{\odot}$ we identify the presence of two distinct modes, with periodicity of 3 minutes and 17 minutes. The two modes survive up to $4R_{\odot}$, but disappears at $6R_{\odot}$. At the farthest radial distance of 12$R_{\odot}$, we observe oscillations with a periodicity of 17.5 minutes, which are essentially the periodicity observed in radial jets and the EUV emissions. The frequency filtering at higher altitudes can also be attributed to the progressively coarser resolution as one moves farther away from the reconnection site.

Similar behavior is observed for different emergence rates, where faster emergence rates result in larger wave amplitudes.

\subsection{Long-lasting oscillation}
\begin{figure}[h]
    \centering
    \hspace*{-1cm}\rotatebox{90}{ $ \hspace{1.8cm} {r/R_{\odot}} $}
    \includegraphics[width=0.48\textwidth]{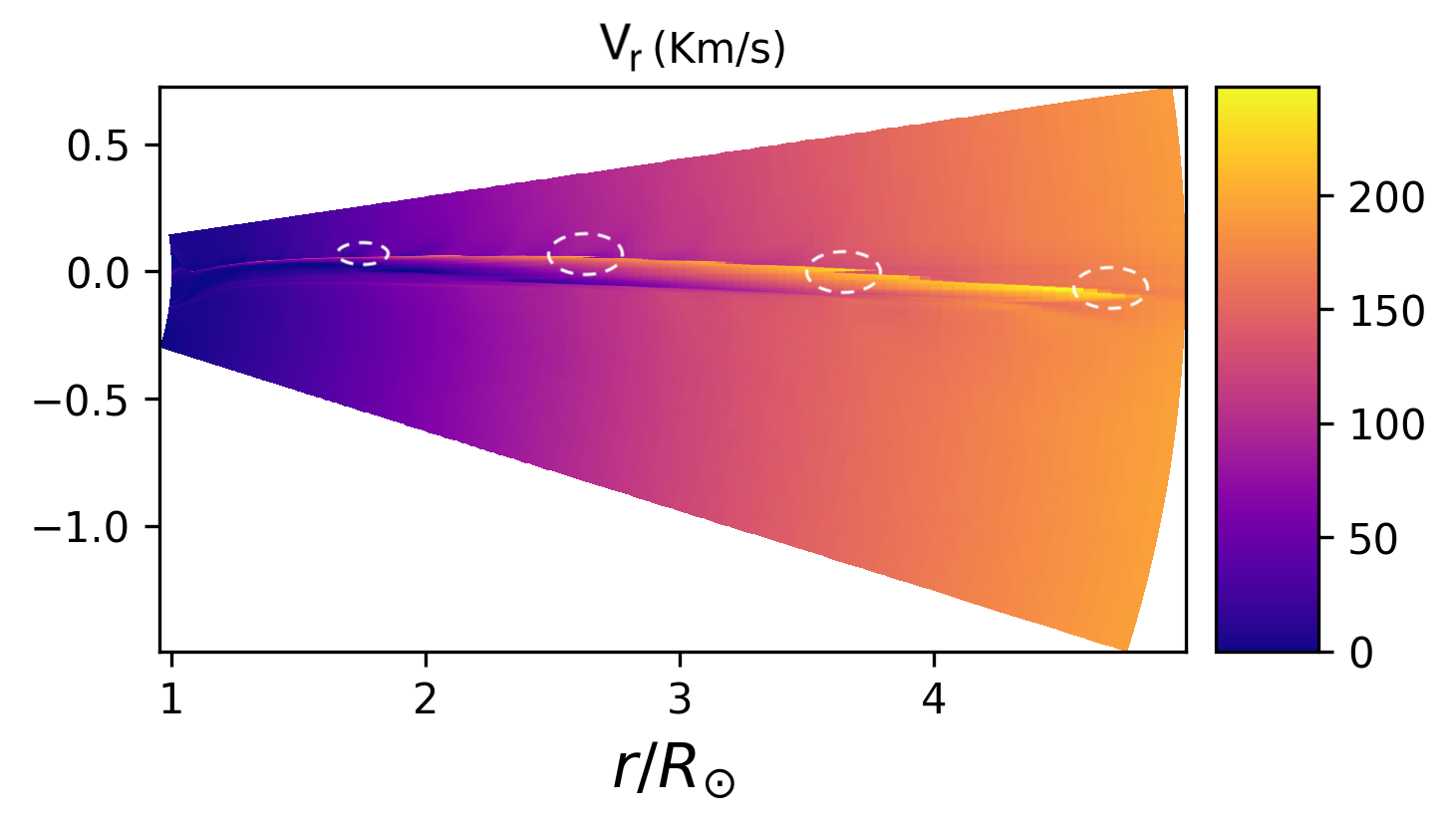}
    \caption{Figure illustrates jet wisp triggered by magnetic reconnection, white circles indicates the wisp signatures}.\hspace*{-1cm}
    \label{fig:Wisp}
\end{figure}

\begin{figure}[h]
    \centering
    \hspace*{-1cm}\rotatebox{90}{ $ \hspace{1.8cm} {r/R_{\odot}} $}
    \includegraphics[width=0.48\textwidth]{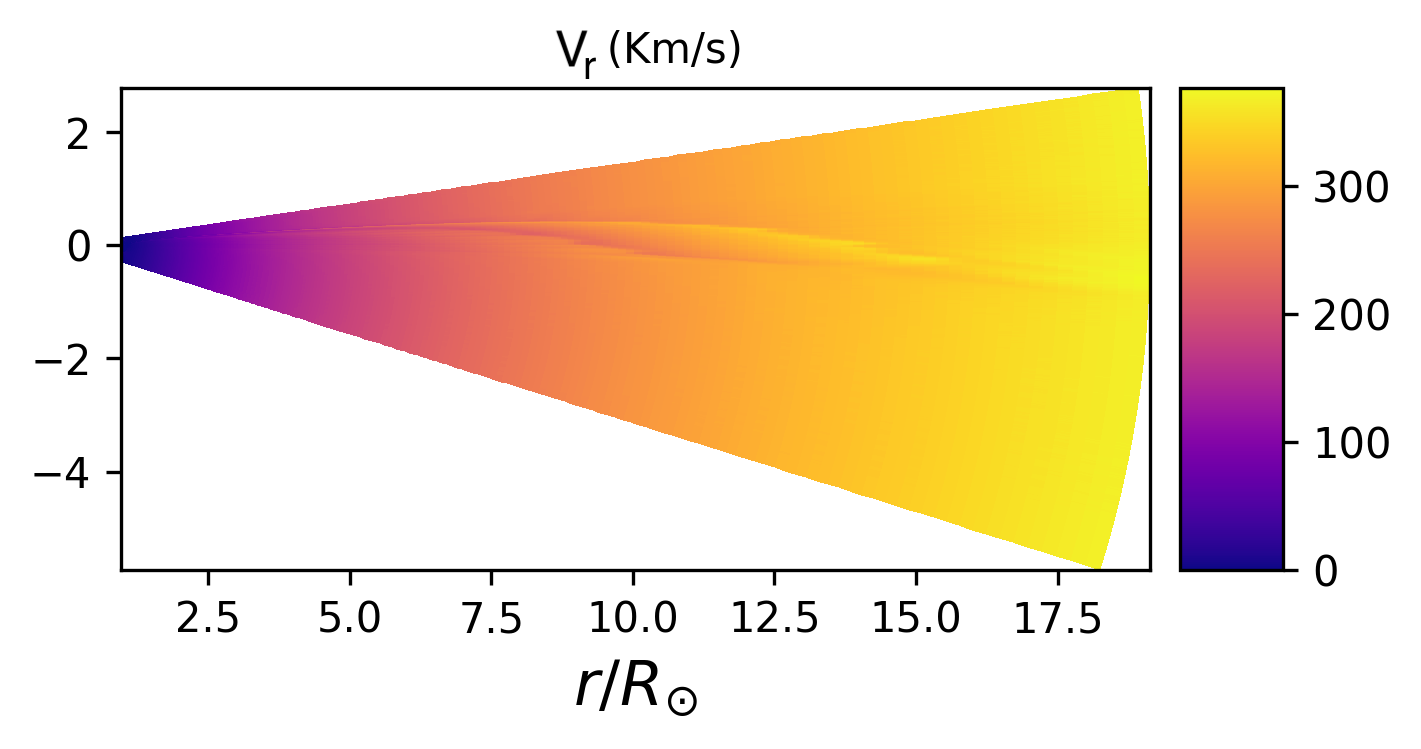}
    \caption{Figure illustrates jet expansion  during current sheet evolution}.\hspace*{-1cm}
    \label{fig:Wiggling}
\end{figure}

Various studies have investigated the transverse motion of coronal jets \citep{1992PASJ...44L.173S,1996ApJ...464.1016C,2007PASJ...59S.771S}.
In our simulations, we observe that the jet is displaced to higher latitudes with a velocity $v_{\theta} \sim 20$km/s comparable with those reported in these past studies. Figure \ref{fig:Wisp} illustrates whip-like motion following the triggered reconnection, the evolution of the current sheet leads to the propagation of velocity spikes characterized by developing fluctuations along the left side of the pseudo-streamer stalk. A similar behavior can be observed in the density perturbation, as shown in Figure \ref{fig:RHO}, where a wisp-like structure is omnipresent with velocity spikes that are initiated by magnetic reconnection within the CS and move away from the stalk as previously reported in observation \citet{2007PASJ...59S.745S}. In particular, these whip-like oscillations consistently align on one side, contributing to the formation of a long-lasting transverse oscillation phenomenon in Figure \ref{fig:Wiggling}. 

The emergence rate has an effect on the jet expansion in the theta direction, this lateral expansion was also reported in coronal observations by \citet{2010ApJ...720..757M} describing expanding jets as "curtain-like spires". Our simulations reveal two driving mechanisms for these transverse motions: from the ideal tearing reconnection process and the expanding motion influenced by magnetic flux emergence.

\section{Discussions}\label{sec:Disc}

In this work, we study the effect of magnetic flux emergence into 2.5D resistive MHD simulations of the solar corona and wind. Our study shows that the emerging process leads naturally to interchange reconnection with the ambient coronal solution and creates jets, or velocity spikes, that then propagated into the solar wind. We observe two main phases of reconnection process. First, shortly after the start of the emergence, the current sheet is created, then lengthens and thins until it reaches the aspect ratio $a/L \propto S^{-1/3}$. Fast reconnection then proceeds through the so-called 'ideal' tearing instability, creating plasmoids that are ejected either towards the Sun and inner boundary condition or towards the fan of the pseudo-streamers created in the corona. The plasmoid propagates at high speed along the CS and appears to hit the stalk of the pseudo-streamer. The plasmoid is eroded and triggers slow magneto-acoustic waves with jets of amplitude up to 200 km/s. Following the completion of the flux emergence phase, we observe a decrease in both the current sheet length and $\alpha$. This suggests that the magnetic field is settling into a more stable state. The current sheet then diffuses and no more bursty reconnection occurs.
The recent study of \citet{2022ApJ...941L..29W}, investigated interchange reconnection with 3D ideal MHD simulations of pseudo-streamers interacting with the solar corona. Although they reach similar conclusions on the creation of Alfvénic structures in the fan of pseudo-streamers, our work complements and differs in some important aspects. First, by precisely controlling the explicit resistivity of the model, we show that the tearing instability is in the ideal regime, which ensures that the reconnection properties should be relatively independent of the Lundquist number and close to the low coronal regime where $S \sim 10^{14}$. Second, while \citet{2022ApJ...941L..29W} triggers reconnection by surface motions, emergence suffices in our case. Finally,  by varying the emergence rate of the bipole, or equivalently the amplitude of the emerging flux, we have shown that the periodicity of the jets matches the periodicity of the EUV emission of the plasma and that it is roughly independent of the emergence rate. This is due to the fact the characteristic Alfvén time $t_A = L/v_A \sim 3 $ minutes, remains unchanged for higher emergence rates and magnetic field amplitudes, as the current sheet lengthens proportionally to the Alfvén speed. The time between each jet is thus the time for the current sheet to be disrupted, i.e., a few $t_A$, or 19 minutes.
These findings are consistent with previous observations reported by \citet{2021ApJ...907....1U} and \citet{2023ApJ...951L..15K}, further supporting the connection between plume oscillations and jet periodicity. Nevertheless, the rate of flux emergence has a significant impact on the observed emission amplitudes in the extreme ultraviolet (EUV) range. Higher flux emergence rates correspond to larger emission amplitudes, as well as higher amplitudes of the velocity spikes.

Several recent studies based on the observations of Parker Solar Probe suggest that switchbacks may  be caused by jetlets originating from small bipoles located at the base of coronal plumes in coronal holes \citep{2021ApJ...919...96F, Bale2021, Shi2022, 2022ApJ...933...21K, 2023Natur.618..252B}. \citet{1995JGR...10023389N} already suggested that the jetlets may also generate microstreams, which are fluctuations in solar wind speed and density observed in polar coronal holes. Although our current results do not directly demonstrate the generation of magnetic reversals through the emergence of bipoles, they do reveal the presence of Alfvénic perturbations that could potentially evolve into magnetic switchbacks. The study of \citep{2022ApJ...941L..29W} shows a somewhat different structure in 3D, with torsional Alfvén waves launched from the pseudo-streamers fan. Yet, no full reversals (or switchbacks) seem able to survive outside of the closed magnetic structures in the simulations. Nonetheless, true switchbacks could be reformed later on as they propagate in  the solar wind, as non-linear developments of seed Alfvén waves, as suggested by \citet{Squire_2020} and \citet{2021ApJ...918...62M}. This emphasizes the need for further observations and MHD simulations to establish a definitive relationship between magnetic switchbacks and interchange reconnection in the chromosphere and transition region beneath plumes.

\section*{Aknowledgements}
The research of BG, VR and APR was funded by the ERC SLOW\_SOURCE (DLV-819189).

The authors are grateful to Kévin Dalmasse, Benoit Lavraud, Peter Wyper, Marco Velli and Nour E. Raouafi for insightful discussions. The authors also thank A. Mignone and the PLUTO development team, on which the numerical work presented in this paper is based. The 2.5D MHD simulations were performed on the Toulouse CALMIP supercomputer and the Jean-Zay supercomputer (IDRIS), through the GENCI HPC allocation grant A0130410293.

This work benefited also from financial support from the Centre National des Études Spatiales (CNES).
\appendix
\counterwithin{figure}{section}
\section{Figures}
\begin{figure}[ht]   
    \centering
    \begin{interactive}{animation}{video.mp4}
    \includegraphics[width=0.8\textwidth]{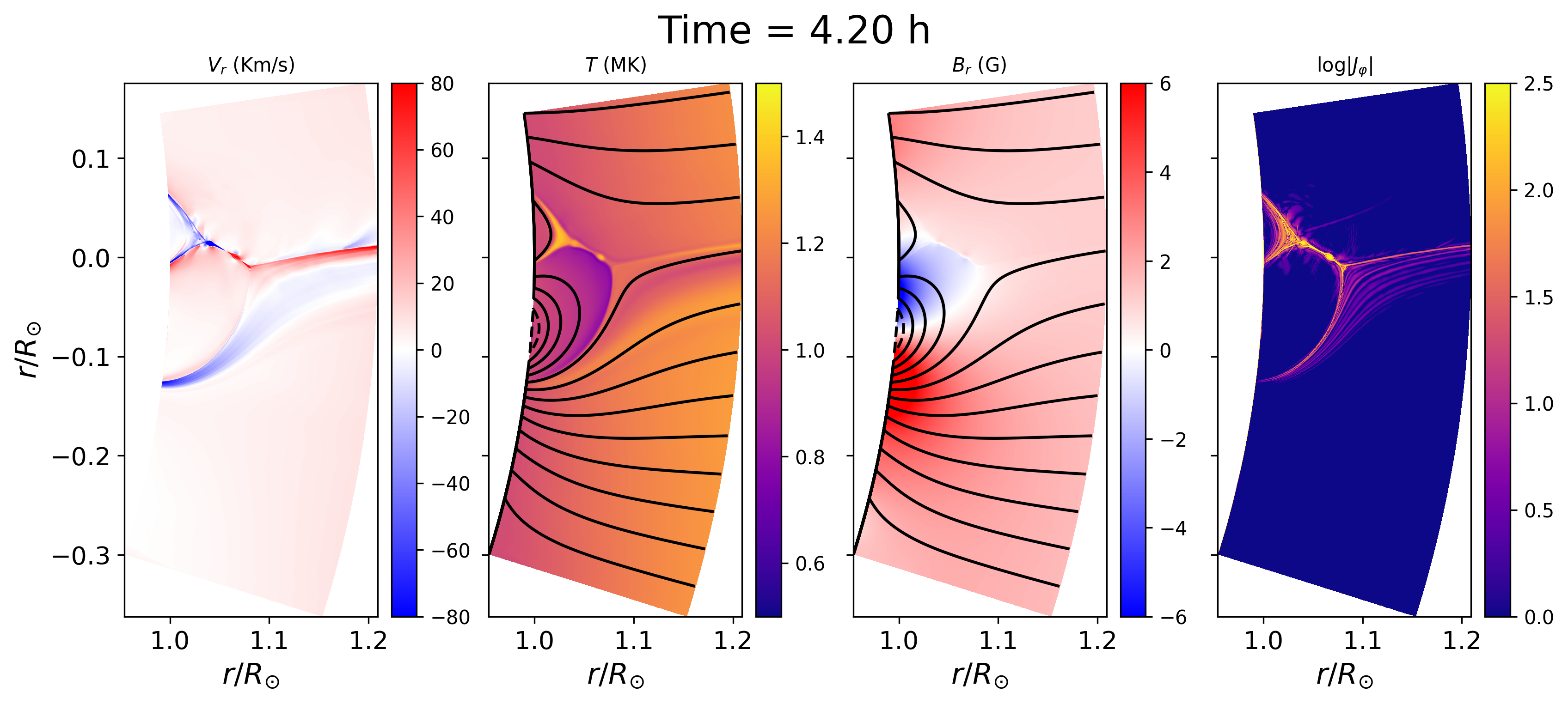}
    \end{interactive}
    \caption{Snapshot of the current sheet during the flux emergence. The four panels display the following variables: radial velocity ($V_r$ in Km/s), temperature (in MK), radial magnetic field ($B_r$ in G), and the logarithm of the out-of-plane current. This figure is available as an animation in the online paper. An animation of this figure is available.}    
    \label{fig:movie}
\end{figure}

Figure \ref{fig:movie} depicts the initiation of the tearing instability for our simulation at an emergence rate of $6$ G/h. The still frame represents the time $t = 4.2$ hours, marking the onset of the tearing instability.

The accompanying animation, associated with Figure \ref{fig:movie}, presents the dynamic evolution of the following variables: radial velocity ($V_r$ in Km/s), temperature (in MK), radial magnetic field ($B_r$ in G), and the logarithm of the out-of-plane current in response to the tearing mode and the multiple instances of instability onset.

\bibliography{sample631}{}
\bibliographystyle{aasjournal}

\end{document}